\documentstyle{article}  
\newlength{\extralineskip}
\newcommand{\beq}{\begin{equation}}
\newcommand{\eeq}{\end{equation}}
\newcommand{\bd}{\begin{displaymath}}
\newcommand{\ed}{\end{displaymath}}

\def\bea{\begin{eqnarray}}
\def\eea{\end{eqnarray}}

\def\ba{\beq\new\begin{array}{c}}
\def\ea{\end{array}\eeq}

\def\inbar{\,\vrule height1.5ex width.4pt depth0pt}
\def\IC{\relax\hbox{$\inbar\kern-.3em{\rm C}$}}
\def\IR{\relax{\rm I\kern-.18em R}}

\def\IN{\relax{\rm I\kern-.15em N}}

\makeatletter
\newdimen\normalarrayskip              
\newdimen\minarrayskip                 
\normalarrayskip\baselineskip
\minarrayskip\jot
\newif\ifold             \oldtrue            \def\new{\oldfalse}
\def\arraymode{\ifold\relax\else\displaystyle\fi} 
\def\@arrayskip{\ifold\baselineskip\z@\lineskip\z@
     \else
     \baselineskip\minarrayskip\lineskip2\minarrayskip\fi}
\def\@arrayclassz{\ifcase \@lastchclass \@acolampacol \or
\@ampacol \or \or \or \@addamp \or
   \@acolampacol \or \@firstampfalse \@acol \fi
\edef\@preamble{\@preamble
  \ifcase \@chnum
     \hfil$\relax\arraymode\@sharp$\hfil
     \or $\relax\arraymode\@sharp$\hfil
     \or \hfil$\relax\arraymode\@sharp$\fi}}
\def\@array[#1]#2{\setbox\@arstrutbox=\hbox{\vrule
     height\arraystretch \ht\strutbox
     depth\arraystretch \dp\strutbox
     width\z@}\@mkpream{#2}\edef\@preamble{\halign \noexpand\@halignto
\bgroup \tabskip\z@ \@arstrut \@preamble \tabskip\z@ \cr}%
\let\@startpbox\@@startpbox \let\@endpbox\@@endpbox
  \if #1t\vtop \else \if#1b\vbox \else \vcenter \fi\fi
  \bgroup \let\par\relax
  \let\@sharp##\let\protect\relax
  \@arrayskip\@preamble}

\begin{document}
\thispagestyle{empty}

\begin{center}
{\huge \bf The Strongly Coupled 't Hooft Model on the Lattice}\\
\vskip 0.3 truein
{\bf F. Berruto, G. Grignani and P. Sodano}
\vskip 0.3truein 
{\it Dipartimento di Fisica and Sezione
I.N.F.N., Universit\`a di Perugia, Via A. Pascoli I-06123 Perugia,
Italy}\\
\vskip 0.3 truein
DFUPG-42-99
\vskip 1.0truein
{\bf Abstract}
\vskip 0.3truein
\end{center}

We study the strong coupling limit of the one-flavor and two-flavor 
massless 't Hooft models, $large-{\cal N}_c$-color $QCD_2$, on a lattice.
We use 
staggered fermions and the Hamiltonian approach to lattice gauge theories. 
We show that the one-flavor model is effectively described by the 
antiferromagnetic Ising model, whose ground state is the vacuum of 
the gauge model in the infinite coupling limit; 
expanding around this ground state we derive a strong coupling expansion 
and compute the lowest lying hadron masses as well as the 
chiral condensate of the gauge theory. 
Our lattice computation well reproduces the results of the continuum theory. 
Baryons are massless in the infinite 
coupling limit; they acquire a mass 
already at the second order in the strong coupling expansion in agreement 
with the Witten argument that baryons are the $QCD$ solitons. 

The spectrum and chiral condensate of the two-flavor model are 
effectively described in terms of observables of
the quantum antiferromagnetic Heisenberg model.
We explicitly write the lowest lying 
hadron masses and chiral condensate in terms of  spin-spin 
correlators on 
the ground state of the spin model. We show that the planar limit 
(${\cal N}_c\longrightarrow \infty$) of the gauge model corresponds to the 
large spin limit ($S\longrightarrow \infty$) of the antiferromagnet and 
compute the hadron mass spectrum in this limit finding that, 
also in this model, the pattern of chiral symmetry breaking of the continuum 
theory is well reproduced on the lattice.

\newpage
\setcounter{page}1

\section{Introduction}
Strong coupling  perturbation theory of QCD is expected to
bring to a simple and correct 
understanding of confinement and chiral symmetry breaking;
furthermore it could provide a very good approximation of 
the low lying hadron spectrum~\cite{wilczek}.

Defining $QCD$ on the lattice provides a gauge invariant ultraviolet cut-off 
convenient to formulate the strong coupling expansion~\cite{wilson}. In the 
strong coupling limit, confinement is explicit, the confining 
string is a stable object and some other qualitative features of the 
spectrum are easily obtained. Although many choices of the strong coupling 
theory produce identical continuum physics, there are strong coupling 
computations which turned out to provide quite accurate 
results~\cite{banks1,shigemitsu, 
banks2, weinstein1, berruto1, steinhardt, berruto2}. 
Strongly coupled lattice gauge theories are intimately related to quantum 
spin systems~\cite{smit, krasnitz, semenoff}; lattice gauge theories with 
staggered fermions~\cite{staggered} exhibit interesting 
similarities with condensed matter systems with lattice fermions. 
For example, it is well known that the quantum spin-$1/2$ Heisenberg antiferromagnet is equivalent to the strong coupling limit of either a $U(1)$ 
or $SU(2)$ lattice gauge theory~\cite{baskaran}. For the gauge group 
$U({\cal N}_c)$, one has equivalence with a spin-${\cal N}_c/2$ Heisenberg 
antiferromagnet~\cite{semenoff}. 
Laughlin~\cite{laughlin} evidenced the relevance of the 
spectral data of gauge theories for the analysis of strongly correlated 
electron systems.              

The idea that quantum antiferromagnetic spin systems are related to quantized gauge theories is very appealing. First of all, 
chiral symmetry 
breaking in the gauge theory corresponds to the 
N\'eel ordering of the quantum 
antiferromagnet~\cite{berruto1, semenoff}. 
Furthermore, at least for some one dimensional gauge theories  it is 
possible~\cite{berruto2, erasmo1} to compute explicitly 
not only the strong coupling massless spectrum in terms of pertinent 
excitations of the exactly solvable~\cite{bethe}
quantum spin-$1/2$ antiferromagnetic Heisenberg chain, 
but also the low-lying masses and chiral condensates in terms of spin 
correlators.
Recently, it has been pointed out that the correspondence 
of gauge theories with quantum antiferromagnets could survive also in the 
weak coupling limit~\cite{weinstein2}. 

Quantum chromodynamics in $1+1$ dimensions, $QCD_2$, 
has played an important 
pedagogical role in our understanding of $QCD$ and in 
motivating approaches to four dimensions. It is a fruitful domain for the 
study of techniques such as bosonization and $1/{\cal N}_c$ expansion. 
In fact, 't Hooft~\cite{thooft} proposed that there is in $QCD_2$ an hidden 
candidate for an expansion parameter: to find it, one should generalize 
$QCD_2$ from three colors and an $SU(3)$ gauge group to 
${\cal N}_c$ colors and an $SU({\cal N}_c)$ gauge group. The hope is then 
to solve the theory in the large-${\cal N}_c$ limit and to show that, 
for ${\cal N}_c=3$, the theory may be qualitatevely and quantitatively 
approximated by the large-${\cal N}_c$ limit. 't Hooft~\cite{thooft2} 
showed that only planar diagrams with the quarks at the edges dominate 
the $1/{\cal N}_c$ perturbative expansion of an $U({\cal N}_c)$ 
gauge theory for ${\cal N}_c\longrightarrow \infty$ with 
$g^2{\cal N}_c$ fixed. 
Even to the lowest order, the exchange of a massless gluon gives rise 
to a linearly attractive potential between a 
quark-antiquark pair in a singlet channel; consequently, 
the two-dimensional model is an interesting nontrivial, confining
quantum field theory and it may be employed as a theoretical laboratory illustrating 
the phenomena occurring in the four-dimensional theory. 
't Hooft solution~\cite{thooft} for the meson spectrum 
is obtained in the weak coupling limit 
\begin{equation}
{\cal N}_c\longrightarrow \infty\   ,\   g^2{\cal N}_c=const.\   ,\   m_q\gg g\sim 1/\sqrt{{\cal N}_c}
\end{equation}
where $m_q$ are the quark masses. In the strong coupling regime $g\gg m_q$ 
the spectrum has been investigated in~\cite{baluni, 
bhattacharya, steinhardt2, amati}. Lattice $QCD_2$ with an $SU(2)$ 
color gauge group has been studied in~\cite{hamer,erasmo1}.
Also on the lattice the large-${\cal N}_c$ limit greatly simplifies the 
analysis of $QCD_2$. 
The simplification may be evidenced through the correspondence of 
the strongly coupled lattice gauge theory with the spin $S={\cal N}_c/2$ 
quantum antiferromagnet (i.e. $S\longrightarrow \infty$ limit ), 
through the mean field theory analysis proposed 
in Ref.~\cite{negele} or through the numerical simulation 
method developed in Ref.~\cite{rebbi}.

In this paper we shall provide an analytical investigation of the one- and 
two-flavor strongly coupled lattice massless 't Hooft models~\cite{thooft} 
using staggered fermions and the hamiltonian approach to lattice gauge 
theories~\cite{kogut2}. We shall investigate both the one-flavor and 
the two-flavor strongly coupled models and 
shall compute not only the meson and baryon 
spectra, but also the pertinent chiral condensates in order to analyse the 
chiral symmetry breaking pattern. 
We shall show that all the physically relevant 
quantities of either the one-flavor or the two-flavor model, at every order 
in the strong coupling expansion, can be computed as vacuum expectation values 
(v.e.v.'s) of spin-spin correlators on the ground state of the antiferromagnetic Ising or Heisenberg model, respectively. 
The two-flavor model is considered here since it is a prototypical 
example for all the multiflavor 't Hooft models and maps $-$ 
at strong coupling $-$ onto a quantum $SU(2)$ Heisenberg antiferromagnet. 
As we shall see,  in the large-${\cal N}_c$ limit 
($S\longrightarrow \infty$), the computation of the mass spectrum and 
pertinent chiral condensate greatly simplifies since the ground state of the 
spin model is frozen in the classical N\'eel configuration.

In $1+1$ dimensions the continuum limit of a staggered fermion 
has the peculiar property of producing exactly one Dirac 
fermion, thus completely avoiding the doubling of fermion species. 
Even though the continuum axial symmetry is broken explicitly by 
staggered fermions, a discrete axial symmetry, corresponding to a 
chiral rotation of $\pi/2$, manifests itself in the lattice theory as a 
translation by one lattice site~\cite{staggered}. 
We shall show that the one-flavor 
't Hooft model spontaneously breaks this discrete axial symmetry 
reproducing all the effects that in the continuum are due to the anomaly, while the two-flavor model exhibits a translationally invariant ground 
state preserving the discrete axial symmetry. The chiral 
anomaly~\cite{jackiw} in the 
two-flavor model is realized on the lattice via the explicit 
breaking of the $U_A(1)$-symmetry induced by the staggered fermions. 
We shall finally extrapolate to the continuum limit both the mass spectrum and the pertinent chiral condensates of the 't Hooft models 
by means of suitable ``Pad\'e approximants''~\cite{pade}.

The one-flavor model in the continuum exhibits a non zero $\langle \bar{\psi}\psi \rangle$ 
chiral condensate~\cite{zhitnitsky}, which we shall compute on the 
lattice up to the fourth order in the strong coupling expansion. We show 
that the answer provided by the lattice computation is in very good 
agreement with the continuum result. 
The two-flavor model exhibits neither an isoscalar 
$\langle \bar{\psi}\psi \rangle$ nor an isovector 
$\langle \bar{\psi} \sigma^a \psi \rangle$ chiral condensate, due to 
the Coleman theorem~\cite{coleman}; on the lattice these chiral 
condensates are zero to all the orders in the strong coupling 
expansion. We shall show that the non-vanishing order parameter for the $U_A(1)$ symmetry 
breaking is given $-$ in analogy with the two-flavor Schwinger model
~\cite{gattringer, berruto2} $-$ by the v.e.v. 
$\langle \overline{\psi^2}_L\overline{\psi^1}_L
\psi^1_R\psi^2_R \rangle$, which we shall compute up to the second order in the 
strong coupling expansion.

In section 2 we review known results of the continuum 
't Hooft model~\cite{thooft}. In section 3 we 
define its Hamiltonian lattice version and we 
identify the lattice counterparts 
of the relevant symmetries of the continuum theory. 

In section 4 we set up the formalism needed for the strong coupling 
analysis of the 't Hooft model. We show that the one-flavor and 
the two-flavor 't Hooft models are effectively described by the 
antiferromagnetic Ising and Heisenberg spin models, respectively.   
 
In section 5 we construct the operators creating the massive meson and baryon excitations. We set up the 
strong coupling expansion and compute the corrections to the energies of the ground state and of the low lying mesons and baryons. 
Subtracting the energy of the ground state from those of the excitations defines the lattice excitation masses. 

In section 6 we compute the $\langle \bar{\psi} \psi \rangle$ chiral condensate of the one-flavor model on the lattice up to the fourth order 
in the strong coupling expansion. We also compute in the two-flavor model, up to the second order in the strong coupling expansion, 
the v.e.v. $\langle \overline{\psi^2}_L\overline{\psi^1}_L \psi^1_R\psi^2_R \rangle$, which is the order parameter for the 
breaking of the $U_A(1)$ symmetry pertinent to this model. 

Section 7 is devoted to the extrapolation of the lattice results to the continuum theory. We shall show that the lattice 
theories, properly extrapolated to the continuum via the use of Pad\'e approximants, well reproduces the parameters of the continuum theory 
already at low orders in the strong coupling expansion.

Section 7 is devoted to some concluding remarks.   

The appendices are introduced to keep the paper self-contained. 
In Appendix A we provide the results of some integrals over the gauge group 
elements needed for the analysis of sections 5 and 6; 
in particular, we compute an integral over six group elements used in the 
strong coupling expansion of the one-flavor model. Appendix B provides some 
technical details useful to derive the mass spectrum and chiral condensates.
          
\section{Large-${\cal N}_{c}$ $QCD_{2}$ }

The large-${\cal N}_{c}$ limit was exploited by 't Hooft to investigate 
two dimensional $U({\cal N}_{c})$-chromodynamics ($QCD_2$), 
where it is possible to sum the planar diagrams.
This simplified version of $QCD$ manifestly
exhibits confinement. 
With ${\cal N}_{f}$ fermion flavors, the 't Hooft model action reads
\begin{equation}
S=\int d^2x \left[\sum_{a=1}^{{\cal N}_{c}}\sum_{\alpha=1}^{{\cal N}_{f}} \overline{\psi}^{\alpha a}(\gamma_{\mu} D^{\mu}+m_{\alpha})\psi^{\alpha}_{a}-
\sum_{a,b=1}^{{\cal N}_{c}}\frac{1}{2}F_{\mu \nu \ b }^{a}F_{\ \ \ a}^{b\mu\nu}\right]\quad ,
\label{tha}
\end{equation}
where, from now on, the first greek letters $\alpha,\beta\ldots$ shall 
denote flavor indices, the latin letters $a,b\ldots$ color 
indices while the greek letters $\mu, \nu,\ldots$ Lorentz indices.  
Furthermore, one has
\begin{eqnarray}
F_{\mu \nu \ b}^{a}&=& \partial_{\mu}A_{b\ \nu}^{a}-\partial_{\nu}A_{a\ \mu}^{b}+ig_c\left[A_{\mu},A_{\nu}\right]_{b}^{a}\quad ,\\
D_{\mu} \psi^{\alpha\ a}&=&\partial_{\mu}\psi^{\alpha\ a}+g_c A_{b\ \mu}^{a}\psi^{\alpha \ b}\quad ,\\
A_{a\ \mu}^{b}(x)&=&(A_{b\ \mu}^{a}(x))^{*}\quad .
\end{eqnarray}

For ${\cal N}_{f}=2$, the action is invariant under the symmetry
\beq
SU({\cal N}_{c})\otimes SU_{L}(2)\otimes SU_{R}(2)\otimes U_{V}(1) \otimes U_{A}(1)\quad .
\label{sym1}
\eeq
The group generators act on the fermion isodoublet to give 
\begin{eqnarray}
SU({\cal N}_{c}) &:& \psi_{\alpha a}(x)\longrightarrow 
(e^{i\Theta_{A}(x)T^{A}})_{ab}\ \psi_{\alpha b}(x)\   , \  
\overline{\psi_{\alpha a}}(x)\longrightarrow \overline{\psi_{\alpha b}}\ (x)
(e^{-i\Theta_{A}(x)T^{A}})_{ba}\quad ,\label{s1} \\
SU_{L}(2) &:& \psi_{\alpha a }(x)\longrightarrow (
e^{i\theta_{B}\frac{\sigma^{B}}{2}P_{L}})_{\alpha \beta}\ \psi_{\beta a }(x)\   , \  
\overline{\psi_{\alpha a}}(x)\longrightarrow \overline{\psi_{\beta b}}\ (x)
(e^{-i\theta_{B}\frac{\sigma^{B}}{2}P_{R}})_{\beta \alpha}\quad ,\label{s11} \\
SU_{R}(2) &:& \psi_{\alpha a}(x)\longrightarrow 
(e^{i\theta_{B}\frac{\sigma^{B}}{2}P_{R}})_{\alpha \beta}\ \psi_{\beta a}(x)\   ,\   
\overline{\psi_{\alpha a}}(x)\longrightarrow \overline{\psi_{\beta a}}(x)\ 
(e^{-i\theta_{B}\frac{\sigma^{B}}{2}P_{L}})_{\beta \alpha}\quad , \\
U_{V}(1) &:& \psi_{\alpha a}(x)\ \longrightarrow 
(e^{i\theta(x){\bf 1}})_{\alpha \beta}\ \psi_{\beta a}(x)\  ,\  
\psi_{\alpha a}^{\dagger}(x)\longrightarrow \psi_{\beta a}^{\dagger}(x)\ 
(e^{-i\theta(x){\bf 1}})_{\beta \alpha} \quad ,\label{s34}\\
U_{A}(1) &:& \psi_{\alpha a}(x)\longrightarrow (e^{i\phi 
\gamma_{5}{\bf 1}})_{\alpha \beta}\ \psi_{\beta a}(x)\  ,\  
\psi_{\alpha a}^{\dagger}(x)\longrightarrow \psi_{\beta a}^{\dagger}(x)\ 
(e^{-i\phi \gamma_{5}{\bf 1}})_{\beta \alpha}\quad ,
\label{s4} 
\end{eqnarray}
where $T^{A}$ are the ${\cal N}_{c}^{2}-1$ generators of the $SU({\cal N}_{c})$ gauge group, 
$\sigma^{B}$ are the Pauli matrices of the flavor group, 
$\Theta_{\alpha}(x)$,$\theta_{\beta}$, $\theta(x)$ and $\phi$ are real coefficients and
\beq
P_{L}=\frac{1}{2}(1-\gamma_{5})\ ,\ P_{R}=\frac{1}{2}(1+\gamma_{5})\quad .
\eeq
At the classical level, the symmetries (\ref{s1}-\ref{s4}) lead to conservation laws for the following currents
\begin{eqnarray}
j^{\mu\ A}(x)&=&\overline{\psi}_{\alpha a}(x)\gamma^{\mu}(T^A)_{ab}\psi_{\alpha b}(x)\quad ,
\label{ca}\\
j_{B}^{\mu}(x)_{R}&=&\overline{\psi}_{\alpha a}(x)\gamma^{\mu}P_{R}
(\frac{\sigma_{B}}{2})_{\alpha \beta}\psi_{\beta a}(x)\quad ,
\label{caa}\\
j_{B}^{\mu}(x)_{L}&=&\overline{\psi}_{\alpha a}(x)\gamma^{\mu}P_{L}
(\frac{\sigma_{B}}{2})_{\alpha \beta}\psi_{\beta a}(x)\quad ,
\label{cb}\\
j^{\mu}(x)&=&\overline{\psi}_{\alpha a}(x)\gamma^{\mu}\psi_{\alpha a}(x)\quad ,
\label{cc}\\
j^{\mu}_{5}(x)&=&\overline{\psi}_{\alpha a}(x)\gamma^{\mu}\gamma ^{5}\psi_{\alpha a}(x)\quad .
\end{eqnarray}

At the quantum level, the vector and axial currents cannot be simultaneously conserved, due to the anomaly phenomenon. If the regularization is gauge invariant, so that the vector currents are conserved, then the axial currents acquire the anomaly~\cite{jackiw}. The axial currents associated to the $SU({\cal N}_{c})$ generators satisfy the anomaly equation for ${\cal N}_{f}$ flavors~\cite{abdalla}
\begin{equation}
\partial_{\mu}j^{5\mu A}+ig\left[A_{\mu},j^{5\mu A}\right]={\cal N}_{f}\frac{g_c^2}{2\pi}\epsilon_{\mu \nu}F^{A\mu \nu}, \quad A=1,\ldots,{\cal N}_{c}^2-1\quad ,
\label{chi1}
\end{equation} 
where $A_{\mu}(x)=A_{\mu}^{C}(x)T^C$, with $T^C,C=1,\ldots,{\cal N}_{c}^{2}-1$ the generators of $SU({\cal N}_{c})$; the abelian axial current satisfies 
the anomaly equation
\begin{equation}
\partial_{\mu}j^{\mu 5}={\cal N}_{f}\frac{g_c^2}{2\pi}
\epsilon_{\mu \nu}F^{\mu \nu} \quad .
\label{chi2}
\end{equation}
The isoscalar $\langle \bar{\psi}\psi \rangle$ and isovector $\langle \bar{\psi}\sigma^{\alpha}\psi \rangle$ 
chiral condensates are zero due to the Coleman theorem~\cite{coleman}; 
in fact, they would break not only the $U_{A}(1)$ symmetry of the action, already broken by the anomaly equation (\ref{chi2}), but also the continuum internal symmetry $SU_{L}(2)\otimes SU_{R}(2)$ down to $SU_{V}(2)$. There is,
 however, the $SU_{L}(2)\otimes SU_{R}(2)$ invariant operator
\begin{equation}
F=\overline{\psi^2}_{L}\overline{\psi^1}_{L}\psi_{R}^{1} \psi_{R}^{2}\quad ,
\label{io}
\end{equation}
 which is non-invariant under the $U_{A}(1)$ symmetry; it can acquire a 
v.e.v. without violating Coleman's theorem~\cite{coleman} 
and consequently may be regarded as a good order parameter for the 
$U_{A}(1)$-breaking.    

If one considers the ${\cal N}_{f}=1$ 't Hooft model, then the action would be invariant under the symmetry group
\beq
SU({\cal N}_{c})\otimes U_{V}(1) \otimes U_{A}(1)\quad ,
\label{sym2}
\eeq 
and, in this case, the group generators action is given only by Eqs.(\ref{s1}),(\ref{s34}),(\ref{s4}), with $\alpha=\beta=1$.
Since the $U_{A}(1)$-symmetry breaking is already realized by the anomaly 
equation (\ref{chi2}), for this model one expects a non zero v.e.v. 
$\langle \overline{\psi} \psi\rangle$. In the weak coupling limit, $i.e.$ for ${\cal N}_{c}\rightarrow \infty$, $g_c^2{\cal N}_{c}=const.$ and the 
fermionic mass $m_{q}\gg g_c\sim 1/\sqrt{{\cal N}_{c}}$, 
Zhitnitsky~\cite{zhitnitsky} computed, taking at the end 
of his computation the limit $m_q\rightarrow 0$, the chiral condensate and found that
\begin{equation}
\langle \overline{\psi}\psi \rangle=-{\cal N}_{c}(\frac{g_c^{2}{\cal N}_{c}}{12\pi})^{\frac{1}{2}}\quad .
\label{zchi}
\end{equation}     

In the $U({\cal N}_{c})$ gauge model the $U_{A}(1)$ chiral symmetry is broken
by the anomaly (\ref{chi2}) and, consequently, there is 
a non-zero $\langle \overline{\psi} \psi\rangle$ without violating Coleman theorem. 
On the contrary, the massless $SU({\cal N}_{c})$ gauge model cannot admit a 
non-zero $\langle \overline{\psi} \psi\rangle$ without violating Coleman 
theorem: if, in fact,  one studies the 
$SU({\cal N}_{c})$ gauge model~\cite{zhitnitsky}, 
a Berezinski-Kosterlitz-Thouless phase transition should be advocated 
to account for a non-vanishing chiral condensate
in the large-${\cal N}_{c}$ limit. 
Since in~\cite{thooft, lcoleman} it was shown that, for 
${\cal N}_{c}\rightarrow \infty$, both $U({\cal N}_{c})$ and $SU({\cal N}_{c})$ describe the same 
physics, one may regard the $U({\cal N}_{c})$ gauge theory as
the version of $QCD_2$ to which Coleman theory does not apply. In this paper 
we shall study the $U({\cal N}_{c})$ gauge model on the lattice in the 
strong coupling limit $g_c\longrightarrow \infty$.

't Hooft~\cite{thooft} was able to sum the planar diagrams of $U({\cal N}_{c})$ $QCD_2$, dominant in the large-${\cal N}_c$ limit, 
with quark and antiquark lines on the boundary and 
derive an effective Bethe-Salpeter equation. 
This reduces to the integral equation
\begin{equation}
\mu^2\phi(x)=(\frac{M_1^2}{x}+\frac{M_2^2}{1-x})\phi(x)-\frac{g_c^{2}}{\pi}P\int_{0}^{1}dy\frac{1}{(x-y)^{2}}\phi(y)\quad ,
\label{inteq}
\end{equation}
where $P$ denotes the principal value integral, $\phi$ is 
defined on the interval $\left[0,1\right]$ and vanishes at the end points of the interval and
\begin{equation}
M_{1,2}=m_{1,2}^{2}-(\frac{g_c^{2}}{\pi})\quad .
\end{equation}
For each eigenvalue of Eq.(\ref{inteq}) there is a bound state of mass $\mu$. Even if Eq.(\ref{inteq}) is not solved analytically, an approximate solution 
yields a linear spectrum of mesonic bound states of quarks given by 
\begin{equation}
\mu_{n}^{2}=g_c^{2}\pi n +(m_1^2+m_2^2 -2\frac{g_c^{2}}{\pi})\ln n+\ldots ,\quad n=0,1,2,\ldots\quad \quad .
\end{equation}
$\mu_{n}$ is the mass of the mesonic bound state.  

Witten~\cite{witten} succeeded in explaining what baryons are like for large ${\cal N}_{c}$. Baryons are a completely antisymmetric state of ${\cal N}_{c}$ quarks and their mass is of order ${\cal N}_{c}$~\cite{witten}. 
In the weak coupling limit, $QCD_2$ exhibits, 
apart from the mesons states, 
additional states whose mass diverges as the inverse of the 
coupling constant: the solitons. Witten, 
firstly, pointed out that baryons are such solitons, since 
their mass diverges with ${\cal N}_{c}$ and $1/ {\cal N}_{c}$ 
is the ``coupling constant'' of the strong interactions. 
In Ref.~\cite{rajeev} it was shown that $QCD$ baryons may be regarded 
as solitons of an effective lagrangian.

\section{$QCD_2$ on the lattice}

We shall use the hamiltonian formulation of lattice gauge theories 
where the time is continuous and the space is a latticized periodic 
chain~\cite{kogut2}. 
The gauge field $U(x)$ associated with the link $\left[x,x+1\right]$ is a group element of $U({\cal N}_{c})$ in the fundamental representation of 
$SU({\cal N}_{c})$ and also carries a representation of $U(1)$. It has the property $U(x,-\hat{i})=U^{\dagger}(x-1,\hat{i})$, where $\hat{i}$ is the 
chain direction vector. 
The electric field $E^{A}_{x}$, associated with the link 
$\left[x,x+1\right]$, satisfies the Lie algebra
\begin{equation}
\left[E^{A}_{x},E^{B}_{y}\right]=if^{ABC}E^{C}_{x}\delta(x-y)\quad ,
\label{lie1}
\end{equation}
and 
\begin{equation}
E(x,-\hat{i})=U^{\dagger}(x-1,\hat{i})E(x-1,\hat{i})U(x-1,\hat{i})\quad ,
\end{equation}
where $E(x,\hat{i})=E^{A}_{x}T^{A}$ with $T^{A}$, $A=0,\ldots,{\cal N}_{c}^{2}-1$, the generators of the Lie algebra of $U({\cal N}_{c})$ obeying 
$[T^{A},T^{B}]=if^{ABC}T^{C}$, $T^{0}=\bf{1}$ the generator of $U(1)$ and $T^{A}$, $A\neq 0$, the generators in the fundamental representation of $SU({\cal N}_{c})$. The electric field generates the left action of the Lie algebra on $U(x)$, $i.e.$, 
\begin{equation}
\left[E^{A}_{x},U(y)\right]=-T^{A}U(x)\delta_{x,y}\quad ,\quad 
\left[E^{A}_{x},U^{\dagger}(y)\right]=U^{\dagger}(x)T^{A}\delta_{x,y}\quad .
\label{lac}
\end{equation}

The Hamiltonian, gauge constraints and non-vanishing (anti-)commutators
of the continuum ${\cal N}_{f}$-flavor 't Hooft model are
\begin{eqnarray}
H=\int dx&&\left[\frac{g_c^2}{2}\sum_{A=0}^{{\cal N}^2_{c}-1}E^A(x)E^A(x)+\sum_{a=1}^{{\cal N}_{c}}\sum_{\alpha=1}^{{\cal N}_{f}}
\psi_{\alpha a}^{ \dagger} (x)\alpha D_{x}\psi_{\alpha a}(x)\right]\ ,\quad\quad
\label{ham1}\\
\partial_x E^{A}(x)+ig_c\left[A^{A}(x),E^{A}(x)\right]&+&\sum_{a,b=1}^{{\cal N}_{c}}\sum_{\alpha=1}^{{\cal N}_{f}} \psi^{\dagger}_{\alpha a}(x)
T^{A}_{ab}\psi_{\alpha b} (x)\sim 0\ ,\ A=1,\ldots, {\cal N}_{c}^{2}-1\ ,\label{ga1}\\
\partial_x E^0(x)&+&\sum_{a=1}^{{\cal N}_{c}}\sum_{\alpha =1}^{{\cal N}_{f}}\psi_{a x}^{\alpha \dagger}\psi_{a x}^{\alpha}\sim 0\quad ,\label{ga11}\\
\left[ A^{A}(x),E^{B}(y)\right]&=&i\delta^{AB}\delta(x-y) ~,
 \left\{\psi_{\alpha a}(x),\psi_{\beta b}^{\dagger}(y)\right\}=\delta_{\alpha \beta}\delta_{ab}\delta(x-y)\ \ .
\label{commu1}
\end{eqnarray}
A lattice Hamiltonian, constraints and (anti-) commutators reducing to Eqs.(\ref{ham1}),(\ref{ga1}),(\ref{ga11}),(\ref{commu1}) 
in the naive continuum limit are 
\begin{eqnarray}
H=\frac{g_L^{2}a}{2}\sum_{x=1}^N\sum_{A=0}^{{\cal N}^2_{c}-1} E_{x}^{A}E_{x}^{A}&-&\frac{it}{2a}\sum_{x=1}^N
\sum_{a,b=1}^{{\cal N}_{c}}\sum_{\alpha=1}^{{\cal N}_{f}} \left(\psi_{a,x+1}^{\alpha \dag}U_{ab}(x)\psi_{b,x}^{\alpha} 
-\psi_{a,x}^{\alpha \dag}U_{ab}^{\dag}(x)\psi_{b,x+1}^{\alpha}\right)\ ,\quad \quad \label{hamilton}\\
{\cal G}^{A}_x&=&E^{A}_x-U^{\dagger}(x-1)E^A_{x-1}U(x-1)+\sum_{a,b=1}^{{\cal N}_{c}}\sum_{\alpha=1}^{{\cal N}_{f}}\psi_{a x}^{\alpha \dagger}
T^{A}_{ab}\psi_{b x}^{\alpha}\sim 0\ ,
\label{gauss1}\\
{\cal G}^{0}_x&=&E^{0}_x-E^{0}_{x-1}+\sum_{a=1}^{{\cal N}_{c}}\sum_{\alpha =1}^{{\cal N}_{f}}\psi_{a x}^{\alpha \dagger}\psi_{a x}^{\alpha}-
{\cal N}_{f}{\cal N}_{c}/2\sim 0\ ,
\label{gauss2}\\
\left[ A^A _x,E^B_y\right]&=&i\delta^{AB}\delta_{x,y}\quad ,\quad \left\{\psi^{\alpha}_{a,x},\psi^{\beta \dagger}_{b,y} \right\}=\delta^{\alpha \beta}\delta_{ab}\delta_{xy}\ .
\end{eqnarray}
The generators of static gauge transformations ${\cal G}^{A}_x$, $A=0,\ldots,{\cal N}_{c}^{2}-1$, obey the Lie algebra
\begin{equation}
\left[{\cal G}^{A}_x,{\cal G}^{B}_y\right]=if^{ABC}{\cal G}^{C}_x\delta_{xy}\quad .
\label{lie2}
\end{equation}
 
The fermion fields are defined on the sites, $x=1,...,N$, 
the gauge and electric fields, $U(x)$ and
 $E^A _{x}$,  on the links $[x; x + 1]$, $N$ is an even integer 
and, when $N$ is finite it is convenient to impose periodic boundary conditions.  When $N$ is finite, the continuum limit is the 
${\cal N}_{f}$-flavor 't Hooft model on a circle.
The coefficient $t$ of the hopping term in Eq.(\ref{hamilton})
plays the role of the lattice light speed. In the naive continuum limit,
$g_L=g_c$ and $t=1$. 

The Hamiltonian and gauge constraints exhibit the discrete symmetries 
\begin{itemize}
\item{}Parity P: 
\begin{equation}
U(x)\longrightarrow U^{\dagger}(-x-1),\ E^A_{x}\longrightarrow -E^A_{-x-1},\ 
\psi_{a,x}^{\alpha}\longrightarrow (-1)^{x}\psi_{a,-x}^{\alpha},\ \psi_{a,x}^{\alpha \dag}\longrightarrow 
(-1)^{x}\psi_{a,-x}^{\alpha \dag}\quad .
\label{par}
\end{equation}

\item{}Discrete axial symmetry $\Gamma$: 
\begin{equation}
U(x)\longrightarrow U(x+1),\ E^A_{x}\longrightarrow E^A_{x+1},\ 
\psi_{a,x}^{\alpha}\longrightarrow \psi_{a,x+1}^{\alpha},\ \psi_{a,x}^{\alpha \dag}\longrightarrow 
\psi_{a,x+1}^{\alpha \dag}\quad .
\label{chir}
\end{equation}

\item{}Charge conjugation C:
\begin{equation}
U(x)\longrightarrow U^{\dagger}(x+1),\  E_{x}\longrightarrow -E_{x+1},\ 
\psi_{a,x}^{\alpha}\longrightarrow \psi^{\alpha \dag}_{a,x+1},\ \psi_{a,x}^{\alpha \dag}\longrightarrow 
\psi_{a,x+1}^{\alpha}\quad .
\label{char}
\end{equation}
\end{itemize}

Since the charge density has to be odd under charge conjugation, 
we defined it in Eq.(\ref{gauss2}) as 
$\rho_x=\sum_{a=1}^{{\cal N}_{c}}\sum_{\alpha =1}^{{\cal N}_{f}}\psi_{a x}^{\alpha \dagger}\psi_{a x}^{\alpha}-
{\cal N}_{f}{\cal N}_{c}/2$. 

\section{The strong coupling limit and the antiferromagnetic Ising and Heisenberg Hamiltonians}

In this section we shall focus on the one- and two-flavor strongly 
coupled 't Hooft models and we shall show 
that they are effectively described by the Ising and Heisenberg 
antiferromagnetic spin models respectively. 
We restrict ourselves to the $SU(2)$-flavor group, a prototypical 
model for all the multiflavor models, for two 
reasons: first, we are interested in the lowest lying states
of the hadron spectrum and heavy quarks are not expected to be 
important in this model; second, we shall show how the 
strong coupling limit of two-flavor $QCD_2$ is 
intimately related to the physical $SU(2)$ antiferromagnet.  

For ${\cal N}_f=1$, the 't Hooft Hamiltonian given in Eq.(\ref{hamilton}), 
rescaled by the factor $g_L^2 a/2$, reads
\begin{equation}
H_t=H_{0}+\epsilon H_{h}
\label{thoh1}
\end{equation}
with 
\begin{eqnarray}
H_{0}&=&\sum_{x=1}^{N}\sum_{A=0}^{{\cal N}_{c}^{2}-1}E_{x}^{A}E_{x}^{A}\quad ,
\label{maxw1}\\
H_{h}&=&-i(R-L)
\label{hopp1}
\end{eqnarray}
and $\epsilon=t^2/g^2_L a^2$. In Eq.(\ref{hopp1}) the 
right $R$ and left $L$ hopping operators are defined ($L=R^{\dagger}$) as 
\begin{equation}
R=\sum_{x=1}^{N}\sum_{a,b=1}^{{\cal N}_{c}}\psi_{a,x+1}^{\dag}
U_{ab}(x)\psi_{b,x}\quad . 
\label{rhopp1}
\end{equation}  

On a periodic chain the commutation relation
\begin{equation}
\left[R,L\right]=0
\label{commu}
\end{equation}
is satisfied. 

We shall consider the strong coupling perturbative expansion where 
the Yang Mills Hamiltonian Eq.(\ref{maxw1}) 
is the unperturbed Hamiltonian and 
the hopping Hamiltonian, Eq.(\ref{hopp1}), is the perturbation. 
 Both terms $H_{0}$ and $H_{h}$ in
 Eq.(\ref{thoh1}) are separately gauge invariant, 
 i.e. $\left[ {\cal G}^{A}_x,H_{0,h}\right]=0$.
Consequently, if one finds a gauge invariant eigenstate of $H_0$, 
perturbations in $H_h$ retain gauge invariance. 

Due to the Gauss law constraints, Eqs.(\ref{gauss1},\ref{gauss2}), 
the lowest energy eigenstate of 
$H_0$ is a color singlet with a density of states of ${\cal N}_c/2$ 
fermions per site. A color singlet at a given site may be 
formed either by leaving the site unoccupied or by putting on it $
{\cal N}_c$ fermions by means of the creation operator   
\begin{equation}
\psi_{a_{1} x}^{\dagger}\ldots \psi_{a_{{\cal N}_{c}}x}^{\dagger}\quad .
\label{cosi1}
\end{equation}

Due to Fermi statistics, one can put at most one singlet per site 
and one can distribute the $N/2$ singlets arbitrarily among the $N$ sites: 
there are then $\left(\begin{array}{c}N\\N/2\end{array}\right)$ 
degenerate ground states. The local fermion number operator
\begin{equation}
\rho_{x}=\sum_{a=1}^{{\cal N}_{c}}\psi_{a x}^{\dagger}\psi_{a x}-{\cal N}_{c}/2
\end{equation}
assumes the value $+{\cal N}_{c}/2$ on occupied sites and $-{\cal N}_{c}/2$ on empty sites. 

First order perturbations to the vacuum energy vanish. 
The ground state degeneracy is removed at the second order 
in the strong coupling expansion \cite{banks2,berruto2}. 
The vacuum energy $-$ at order $\epsilon^2$ $-$ reads
\begin{equation}
E_{0}^{(2)}=\langle H_{h}^{\dagger}\frac{\Pi_s}{E_{0}^{(0)}-H_{0}}H_{h} \rangle\quad ,
\label{soen}
\end{equation}
where the expectation values are defined on the degenerate subspace and $\Pi$ is the operator projecting on a set orthogonal to the degenerate ground 
states. The following commutator holds effectively in the degenerate ground state subspace
\begin{equation}
\left[ H_{0},H_{h}\right]=C_{2}^{f}({\cal N}_{c})H_{h}\quad ,
\label{c2}
\end{equation}
where $C_2^f({\cal N}_c)={\cal N}_c/2$ is the quadratic Casimir of the fundamental representation of $U({\cal N}_c)$.
Using Eq.(\ref{c2}), from Eq.(\ref{soen}) one gets
\begin{equation}
E_{0}^{(2)}=-\frac{2}{C_{2}^{f}({\cal N}_{c})}\langle RL \rangle \quad .
\label{seon2}
\end{equation}
The vacuum expectation values $\langle \ ,\ \rangle$ are the inner products in the full Hilbert space of the model, $i.e.$ 
$\langle \ ,\ \rangle=\prod_{x}\int dU_{x} (\ ,\ )$, where $dU$ is the Haar measure on the gauge group manifold and $(\ ,\ )$ is the fermion Fock space inner product. 
Using the integrals over the group elements, given in 
Eqs.(\ref{i1},\ref{i2}) of Appendix A, the combination $RL$ 
may be written in terms of an Ising spin Hamiltonian; in fact, 
since products of $L_{x}$ and $R_{y}$ at different points have vanishing expectation values, due to Eq.(\ref{i2}), 
one can rewrite Eq.(\ref{seon2}) as 
\begin{equation}
E_{0}^{(2)}=\frac{2}{C_{2}^{f}({\cal N}_{c}){\cal N}_{c}}\ \langle H_I \rangle\quad ,
\end{equation}
with
\begin{equation}
H_I=\sum_{x=1}^{N}(\rho(x)\rho(x+1)-\frac{1}{4}{\cal N}_{c}^{2})\quad .
\label{ising}
\end{equation}
The Hamiltonian given in Eq.(\ref{ising}) is an antiferromagnetic Ising Hamiltonian in the space of pure fermion states where 
$\rho_{x}=\pm{\cal N}_{c}/2$ and 
$\sum_{x=1}^{N}\rho_x=0$. The Hamiltonian has two degenerate ground states characterized by a fermion distribution 
$\rho_{x}=\pm{\cal N}_{c}/2 (-1)^{x}$, which do not mix at any finite 
order of the strong coupling expansion. Consequently, as in the 
one-flavor Schwinger model~\cite{berruto1}, one should carry out
non degenerate perturbation theory around one of the two ground states, 
$|g.s.\rangle $.  

Consider now the two-flavor lattice 't Hooft  model. 
The Hamiltonian of the model is given by Eq.(\ref{thoh1}), 
where, now, in $H_h$
 the right $R$ and left $L$ hopping operators are d
efined ($L=R^{\dagger}$) as 
\begin{equation}
R=\sum_{x=1}^{N}\sum_{\alpha=1}^{2}\sum_{a,b=1}^{{\cal N}_{c}}\psi_{a,x+1}^{\alpha \dag}U_{ab}(x)\psi_{b,x}^{\alpha}\quad ; 
\label{rhopp}
\end{equation}      
on a periodic chain the commutation relation, 
given in Eq.(\ref{commu}), holds. 

We shall perform a strong coupling expansion paralleling what we did 
for the one-flavor 't Hooft model. 
Again, one has to find the lowest energy gauge invariant eigenstates 
of $H_{0}$ at half-filling, 
with a density of states of ${\cal N}_{f}{\cal N}_{c}/2={\cal N}_c$ 
fermions per site.    
The empty vacuum $|0\rangle$ is a singlet of the algebra given in 
Eq.(\ref{lie1}) and has no fermions 
$\psi_{a x}^{\alpha}|0\rangle=0$, for every $x,a,\alpha$. By acting on $|0\rangle$ with the operator 
\begin{equation}
\psi_{a_1 x}^{\alpha_{1}\dagger}\ldots
\psi_{a_{{\cal N}_{c}}x}^{\alpha_{{\cal N}_{c}} \dagger}\quad ,
\label{cosi}
\end{equation}   
where the greek indices $\alpha_1,\ldots,\alpha_{{\cal N}_c}$ take the values 1,2, one creates a color singlet on the site $x$. Since the operator (\ref{cosi}) is symmetric in the flavor indices, it carries an irreducible representation 
of $SU({\cal N}_{f})$ with a Young tableau with ${\cal N}_{c}$ columns and one row. Due to Fermi statistics there are at most ${\cal N}_{f}$ singlets on a given site; the allowed representations of the flavor $SU({\cal N}_{f})$ algebra on one site are the empty singlet and those with Young tableaux of 
${\cal N}_{c}$ columns and $1,\ldots,{\cal N}_{f}$ rows distinguished by the fermion numbers $\rho_x={\cal N}_{c}(\nu -{\cal N}_{f}/2)$, $\nu=0,1,\ldots,{\cal N}_{f}$, respectively. The states with $N{\cal N}_{f}/2$ color singlets are degenerate ground states of the Hamiltonian $H_{0}$ with zero 
eigenvalue $E_{0}^{(0)}=0$ and are labelled by the fermion density $\rho_x$ and the vector in the corresponding $SU({\cal N}_{f})$ representation at each site. 

As in the one-flavor model, 
the ground state degeneracy is resolved by diagonalizing 
the first non-trivial order in perturbation theory. 
that is the second order Due to Eq.(\ref{i1}) the first order vanishes
so that the first non-vanishing contribution comes from the second order. 
At order $\epsilon^2$ the vacuum energy is given by Eq.(\ref{soen}); 
using the commutator given in Eq.(\ref{c2}) and from Eq.(\ref{soen}),
 one gets Eq.(\ref{seon2}). 
By introducing the Schwinger spin operators for the spin-${\cal N}_{c}/2$ $SU(2)$ algebra in the representation 
with Young tableau of one row and ${\cal N}_{c}$ columns, 
one writes  
\begin{equation}
\vec{S}_{x}=\sum_{a=1}^{{\cal N}_{c}}\sum_{\alpha,\beta=1}^{2} \psi_{ax}^{\alpha \dagger}\frac{\vec{\sigma}_{\alpha \beta}}{2}\psi_{a x}^{\beta}\quad .
\label{scsp}
\end{equation}
Taking into account that products of $L_{x}$ and $R_{y}$ at different points have vanishing expectation values, due to Eq.(\ref{i2}), 
one can rewrite Eq.(\ref{seon2}) as
\begin{equation}
E_{0}^{(2)}=\frac{4}{C_{2}^{f}({\cal N}_{c}){\cal N}_{c}}
\langle H_{J} \rangle \quad .
\label{seon3}
\end{equation}    
In Eq.(\ref{seon3}) $H_J$ is the Heisenberg Hamiltonian given by
\begin{equation}
H_{J}=\sum_{x=1}^{N}(\vec{S}_{x}\cdot\vec{S}_{x+1}-S^2)\quad ,\quad S=\frac{{\cal N}_{c}}{2}\ .
\label{hhe}
\end{equation}
The ground state of $H_{J}$, $|g.s.\rangle$, singles out the correct vacuum, on which to perform the strong coupling expansion. 
The correct vacuum is a translationally invariant spin singlet of the spin-${\cal N}_{c}/2$ $SU(2)$ algebra. 

Let us consider the explicit example of the $U(3)$ gauge group. 
The spin-$3/2$ operators read
\begin{eqnarray}
S^{+}_{x}&=&\sum_{a=1}^{3}\psi_{ax}^{1\dagger}\psi_{ax}^{2}\quad ,\\
S^{-}_{x}&=&\sum_{a=1}^{3}\psi_{ax}^{2\dagger}\psi_{ax}^{1}\quad ,\\
S^{3}_{x}&=&\frac{1}{2}\sum_{a=1}^{3}(n_{ax}^{1}-n_{ax}^{2})\quad .
\end{eqnarray}
The ground state of an $N$ site spin-$3/2$ chain is a spin singlet built as 
a linear combination of states with on every site a configuration given by
\begin{eqnarray}
|S^{3}=3/2\rangle&=&|\begin{array}{c}u \\ u \\ u \end{array}\rangle \quad , \label{ss1}\\
|S^{3}=1/2\rangle&=&\frac{1}{\sqrt{3}}\left(|\begin{array}{c}u \\ u \\ d \end{array}\rangle + |\begin{array}{c}u \\ d \\ u \end{array}\rangle +
|\begin{array}{c}d \\ u \\ u \end{array}\rangle \right)\quad , \label{ss2}\\
|S^{3}=-1/2\rangle&=&\frac{1}{\sqrt{3}}\left(|\begin{array}{c}u \\ d \\ d \end{array}\rangle + |\begin{array}{c}d \\ u \\ d \end{array}\rangle +
|\begin{array}{c}d \\ d \\ u \end{array}\rangle \right)\quad ,\label{ss3}\\
|S^{3}=-3/2\rangle&=&|\begin{array}{c}d \\ d \\ d \end{array}\rangle
\quad .
\label{ss4}
\end{eqnarray}
In Eqs.(\ref{ss1}-\ref{ss4}) one labels a flavor 1 particle 
with a $u$ and a flavor 2 particle with a $d$ and each row represents 
a color index. 
From this simple example one can picture the complexity of 
the ground state for a spin-${\cal N}_c/2$ chain. The ground state of a 
spin-$1/2$ chain is described in~\cite{berruto2,berruto3}.

Even if the spin-$1/2$ antiferromagnetic Heisenberg chain is an integrable system~\cite{bethe}, a quantum spin-${\cal N}_{c}/2$ antiferromagnet is not 
so easy to handle. However, also on the lattice things greatly 
simplify in the large-${\cal N}_{c}$ limit, since $S={\cal N}_c/2$. 
For $S\longrightarrow \infty$ the quantum nature of a magnetic system can be forgotten, and one may regard $\vec{S}$ as a classical vector in the three 
dimensional space, since the spin commutator is much smaller than the square of the spin variables
\begin{equation}
\left[ S^{a},S^{b} \right]=i\epsilon^{abc}S^{c}=O(S)\ll O(S^{2})\quad .
\label{clco}
\end{equation}
For $S\longrightarrow \infty$, $i.e.$ in the classical limit, the 
ground state of the Heisenberg antiferromagnet is frozen in the 
classical N\'eel state, characterized by neighbouring spins antiparallel
\begin{equation}
|N\rangle=|\ \uparrow \downarrow \uparrow \downarrow \uparrow \downarrow \uparrow \downarrow \uparrow \downarrow 
\uparrow \downarrow \uparrow \downarrow \ldots\ \rangle \quad .
\label{neel}
\end{equation}
As we shall see in the next sections, the use of the ground state
(\ref{neel}) to
compute the spin correlators needed to evaluate 
the mass spectrum and chiral condensate of the gauge model, 
is a remarkable simplification. It amounts to compute the spin correlators
at the zeroth order of the spin wave theory. 
Spin-wave theory~\cite{anderson} fails in one dimension due to the infrared 
divergences signaling the destabilization of the N\'eel vacuum by quantum 
fluctuations, in agreement with the Coleman and Mermin-Wagner 
theorems~\cite{coleman}. 
Of course, this is true for a spin-$1/2$ antiferromagnet chain: in fact, 
if one 
uses the spin wave theory to compute spin-spin correlators for a 
spin-$1/2$ antiferromagnetic chain, one gets wrong answers for large 
distances 
since quantum disorder dominates the large scales  and the N\'eel 
state cannot pick up the main features of the quantum ground 
state~\cite{berruto4}. On the contrary, if one uses the spin wave theory 
to analyse an antiferromagnet with really large spin, then the 
predictions of spin wave theory are quite accurate, and become exact 
at $S=\infty$, where, due to Eq.(\ref{clco}), the spin operators 
become spin vectors.

\section{The hadron spectrum}

In this section we shall investigate the meson and baryon spectrum for 
the one and the two-flavor strongly coupled lattice 
$QCD_{2}$ in the large-${\cal N}_{c}$ limit. For both models, 
massive mesons are created by acting on the ground state of the spin Hamiltonian with the lattice currents of the gauge model generating fermion transport 
besides spin flipping, while baryons are created by acting on the spin 
models vacua with suitable on site color singlet operators. 
Baryons are massless at zeroth order in the strong coupling expansion 
and may be regarded as solitons or Polyakov-'t Hooft monopoles, in agreement with the Witten picture~\cite{witten}. 
The soliton mass should be proportional to the inverse of the coupling 
constant: in fact, it diverges in the weak coupling limit, while at the 
second order in the strong coupling expansion it is non-zero and 
proportional to ${\cal N}_{c}$. Furthermore, in close analogy with the two flavor Schwinger model~\cite{berruto2} 
the two-flavor model exhibits also other massless excitations at finite 
${\cal N}_c$, involving only spin flipping in the strongly coupled ground 
state: they are the spinons of the antiferromagnetic spin chain. 
At finite ${\cal N}_c$, according  to the Haldane conjecture~\cite{haldane}, 
the spinons are gapless only 
for chains of half-integer spin, $i.e.$ for  ${\cal N}_{c}$ odd, 
while integer spin chains exhibit a gapped spectrum. 
At ${\cal N}_c=\infty$, we expect that the massless meson excitations 
are given by collective coherent fluctuations on the classical N\'eel ground 
state. 

Let us now investigate the one-flavor 't Hooft model meson spectrum in the 
strong coupling limit. If one evaluates the ground state energy up to the 
fourth order in the strong coupling expansion, one gets
\begin{equation}
E_{g.s.}=\frac{g^2_La}{2}(E_{g.s.}^{(0)}+\epsilon^2E_{g.s.}^{(2)}+\epsilon^4E_{g.s.}^{(4)})
\end{equation}
with
\begin{equation}
E_{g.s.}^{(0)}=\langle g.s.|H_{0}|g.s. \rangle=0\quad ,
\end{equation}
\begin{equation}
E_{g.s.}^{(2)}=\langle g.s.|H_{h}^{\dagger}\frac{\Pi_{g.s.}}{E_{g.s.}^{(0)}-H_{0}}H_{h}|g.s. \rangle=-2{\cal N}_cN\quad ,
\end{equation} 
\begin{eqnarray}
E_{g.s.}^{(4)}&=&\langle g.s.|H_{h}^{\dagger}\frac{\Pi_{g.s.}}{E_{g.s.}^{(0)}-H_{0}}H_{h}^{\dagger}\frac{\Pi_{g.s.}}{E_{g.s.}^{(0)}-H_{0}}
H_{h}\frac{\Pi_{g.s.}}{E_{g.s.}^{(0)}-H_{0}}H_{h}|g.s. \rangle\nonumber\\
& &-\langle g.s.|H_{h}^{\dagger}\frac{\Pi_{g.s.}}{E_{g.s.}^{(0)}-H_{0}}H_{h}|g.s. \rangle 
\langle g.s.|H_{h}^{\dagger}\frac{\Pi_{g.s.}}{(E_{g.s.}^{(0)}-H_{0})^2}H_{h}|g.s. \rangle=16 {\cal N}_{c}N\quad .
\label{egs4}
\end{eqnarray}
$\Pi_{g.s.}$ is the projection operator projecting on states 
orthogonal to $|g.s.\rangle$. If one rescales the coupling constant as 
\begin{equation}
g^2_L\longrightarrow g^2_L{\cal N}_{c}\quad ,
\label{gres}
\end{equation}
the strong coupling expansion parameter becomes
\begin{equation}
\epsilon\longrightarrow \frac{\epsilon}{{\cal N}_{c}}\quad .
\label{eres}
\end{equation}
The rescaling (\ref{gres}) is needed since the meson masses are proportional 
to ${\cal N}_c$ to every order in the strong 
coupling expansion. This infinity is absorbed in the coupling constant 
which provides the right dimensions to the strong coupling series. 

Gauge invariant massive excitations are created by the operators 
\begin{equation}
\hat{O}_n(x)=\sum_{a,b=1}^{{\cal N}_c}\psi^{\dagger}_{a\ x+n}\left(\prod_{j=x}^{x+n}U(j)\right)_{ab}\psi_{b\ x}
\label{hn}
\end{equation}
and $-$ at zeroth order in the strong coupling expansion $-$ have energy 
\begin{equation}
E_n=\frac{g_L^2}{2}C_2^f({\cal N}_c)na 
\end{equation}
greater than the ground state energy. 
Depending on if $n$ is positive or negative, the operator 
$\hat{O}_n(x)$ (\ref{hn}) is a right or left hopping operator, which 
destroys a particle 
in $x$ and creates a particle in $x\pm|n|$. By applying 
the operators $\hat{O}_n=\sum_{x=1}^N\hat{O}_n(x)$ on 
the ground state $|g.s.\rangle$, one creates 
a meson at zero momentum and $-$ for each $n$ $-$  
the energy is that of a quark-antiquark pair bound by a colour flux of 
length $na$.  

In the following we shall turn our attention to the 
lowest lying mesons $-$ which one gets for $n=1$ $-$ and compare 
the results of our computation with those obtained in the continuum in 
Ref.~\cite{bhattacharya}. Of course, the theory is confining 
since, due to gauge invariance, to a single quark should be attached a line 
of color flux which goes from the quark to infinity; this is not allowed in a finite volume with periodic boundary conditions and on the infinite 
lattice where the string of color flux has an infinite energy. 
The lowest lying excitations are degenerate 
and are created by the right and left hopping operators $R$ and $L$. 
When these operators act on the vacuum $|g.s.\rangle$, they create the two 
degenerate states 
\begin{eqnarray}
|R\rangle&=&R|g.s.\rangle\quad ,\\ 
|L\rangle&=&L|g.s.\rangle\quad .
\end{eqnarray}
The degeneracy is removed at the second order in the strong coupling 
expansion and, at zero momentum, the pseudoscalar and scalar excitations which 
separate in energy are given by
\begin{eqnarray}
|P\rangle&=&(R+L)|g.s.\rangle=\sum_{x=1}^Nj^1(x)|g.s.\rangle \quad ,\label{pse}\\
|S\rangle&=&(R-L)|g.s.\rangle=\sum_{x=1}^Nj^5(x)|g.s.\rangle \quad .\label{se}
\end{eqnarray} 
The energy of the state $|P\rangle$, at the fourth order in $\epsilon$, 
is given by
\begin{equation}
E_{P}=\frac{g^2_La}{2}(E_{P}^{(0)}+\epsilon^2E_{P}^{(2)}
+\epsilon^4E_{P}^{(4)})
\label{mp}
\end{equation}
with
\begin{equation}
E_{P}^{(0)}=\langle P|H_{0}|P \rangle=\frac{1}{2}\quad ,
\label{ep1}
\end{equation}
\begin{equation}
E_{P}^{(2)}=\langle P|H_{h}^{\dagger}\frac{\Pi_{P}}{E_{P}^{(0)}-H_{0}}H_{h}|P \rangle=-2{\cal N}_{c}N-4
\label{ep2}
\end{equation} 
and
\begin{eqnarray}
E_{P}^{(4)}&=&\langle P|H_{h}^{\dagger}\frac{\Pi_{P}}{E_{P}^{(0)}-H_{0}}H_{h}^{\dagger}\frac{\Pi_{P}}{E_{P}^{(0)}-H_{0}}
H_{h}\frac{\Pi_{P}}{E_{P}^{(0)}-H_{0}}H_{h}|P\rangle\nonumber\\
& &-\langle P|H_{h}^{\dagger}\frac{\Pi_{P}}{E_{P}^{(0)}-H_{0}}H_{h}|P \rangle 
\langle P|H_{h}^{\dagger}\frac{\Pi_{P}}{(E_{P}^{(0)}-H_{0})^2}H_{h}|P \rangle=16 {\cal N}_{c}N+80\quad .
\label{ep4}
\end{eqnarray}
$\Pi_{P}$ is the projection operator projecting on states  orthogonal to 
$|P\rangle$. To compute the energy at the second and fourth order 
in the strong coupling expansion given by Eqs.(\ref{ep2},\ref{ep4}), one 
uses the 
integrals over the gauge group elements reported in 
Eqs.(\ref{i3},\ref{i4}) of Appendix A. 
Analogously, one can compute the energy of the scalar meson up to the fourth order in the strong coupling expansion; one gets
\begin{equation}
E_{S}=\frac{g^2_La}{2}(E_{S}^{(0)}+\epsilon^2E_{S}^{(2)}+\epsilon^4E_{S}^{(4)})
\label{ms}
\end{equation}
with
\begin{eqnarray}
E_{S}^{(0)}&=&\frac{1}{2}\quad ,\\
E_{S}^{(2)}&=&-2{\cal N}_{c}N+4\quad ,
\label{es2}\\
E_{S}^{(4)}&=&16 {\cal N}_{c}N+80\quad .
\label{es4}
\end{eqnarray}
In Eqs.(\ref{es2},\ref{es4}), have been used
again the integrals over the gauge group elements reported in Appendix A. 

The meson masses are given by
\begin{eqnarray}
m_{P}&=&\frac{g_L^2a}{2}(E_{P}-E_{g.s.})=\frac{g_L^2a}{2}(\frac{1}{2}-4\epsilon^2+80\epsilon^4)\quad ,\label{mpse}\\
m_{S}&=&\frac{g_L^2a}{2}(E_{S}-E_{g.s.})=\frac{g_L^2a}{2}(\frac{1}{2}+4\epsilon^2+80\epsilon^4)\quad .\label{msca}
\end{eqnarray}
In Eqs.(\ref{mpse},\ref{msca}), the ($N$-dependent) ground state energy terms appearing in $E_{P}^{(2)}$ and 
$E_{S}^{(2)}$ cancel and what is left are only $N$ independent terms. This is a good check of our computation, 
since the mass an intensive quantity.

The one-flavor model spectrum exhibits only one kind of baryon 
which is created at 
zero momentum by acting on the ground state with the color singlet operator 
$B^{\dagger}$
\begin{equation}
|B\rangle=B^{\dagger}|g.s.\rangle=\sum_{x=1}^{N}B^{\dagger}_x|g.s.\rangle= 
\sum_{x=1}^{N}\psi_{1x}^{\dagger}\psi_{2x}^{\dagger}\ldots\psi_{{\cal N}_{c}x}^{\dagger}|g.s.\rangle\quad .
\end{equation}
At zeroth order in the strong coupling expansion, the baryon is massless, since the creation operator $B^{\dagger}$ does not 
contain any color flux
\begin{equation}
H_{0}|B\rangle=0\quad .
\end{equation}
At the second order, since $H_{h}$ and $B^{\dagger}$ do not commute, baryons acquire a mass given by
\begin{eqnarray}
E_{B}^{(2)}&=&\frac{1}{\langle B|B\rangle}\langle B|H_{h}^{\dagger}\frac{\Pi_{B}}{E_{B}^{(0)}-H_{0}}H_{h}|B\rangle=-2{\cal N}_{c}N+2\quad ,\quad \\
m_{B}^{(2)}&=&g^2_La\quad .
\end{eqnarray} 
What one finds in the strong coupling limit is in agreement with the Witten conjecture that baryons are solitons or 
Polyakov-'t Hooft monopoles. In fact, baryons are massless in the infinite coupling limit, but start acquiring a mass 
already at the second order in the strong coupling expansion and their mass is proportional to ${\cal N}_c$ due to Eq.(\ref{gres}).

Let us now turn our attention to the two-flavor $QCD_{2}$ spectrum. 
The ground state energy, up to the second order in the strong coupling 
expansion, reads
\begin{equation}
E_{g.s.}=\frac{g_L^2a}{2}\left( E_{g.s.}^{(0)}+\epsilon^2 E_{g.s.}^{(2)}\right) 
\end{equation}
with
\begin{eqnarray}
E_{g.s.}^{(0)}&=&\langle g.s.|H_0|g.s.\rangle =0\quad , 
\label{qgs1}\\
E_{g.s.}^{(2)}&=&\langle g.s.|H_h^{\dagger}\frac{\Pi_{g.s.}}{E_{g.s.}^{(0)}-H_0}H_h|g.s.\rangle =\frac{4}{C_2^f({\cal N}_c)}
\langle g.s. |H_J|g.s.\rangle\quad .
\label{qgs2}
\end{eqnarray}

Two mesons are now created using the spatial components of the 
vector $j^{1}(x),$ Eq.(\ref{cc}), and the isovector $j_{B}^{1}(x)$ 
given in Eqs.(\ref{caa},\ref{cb}) currents, respectively. 
They are a pseudoscalar isosinglet and a pseudoscalar isotriplet. 
While the vector current, given in Eq.(\ref{cc}), is manifestly associated 
with a Bose field via abelian bosonization~\cite{coleman3}, the 
isovector current, given in Eqs.(\ref{caa},\ref{cb}),  
has a complicated nonlinear and nonlocal expression in terms of Bose fields. 
A more symmetrical treatment of the bosonized form of the isotriplet currents 
is available within the framework of non-abelian bosonization~\cite{gepner}. 
Even if, in the continuum theory, the states created by applying the 
isovector currents to the vacuum are not simple particle states, 
on the lattice one is able to compute their energy using a strong coupling 
expansion.  

The lattice operators which, acting on the spin chain ground state $|g.s.\rangle$, 
create the meson states at zero momentum  with the correct quantum numbers, 
read
\begin{eqnarray}
S&=&R+L=\sum_{x=1}^{N}j^{1}(x)\quad ,\\
T_{+}&=&(T_{-})^{\dagger}=R^{(12)}+L^{(12)}=\sum_{x=1}^{N}j_{+}^{1}(x)\quad ,
\label{ta}\\
T_{0}&=&\frac{1}{\sqrt{2}}(R^{(11)}+L^{(11)}-R^{(22)}-L^{(22)})=
\sum_{x=1}^{N}j_{3}^{1}(x) \quad .
\label{t00}
\end{eqnarray}
$R^{(\alpha\beta)}$ and $L^{(\alpha\beta)}$ in Eqs.(\ref{ta},\ref{t00}) are the 
right and left flavor changing 
hopping operators ($L^{(\alpha\beta)}=(R^{(\alpha\beta)})^{\dagger}$) 
given by
$$
R^{(\alpha\beta)}=\sum_{x=1}^{N}\sum_{a,b=1}^{{\cal N}_{c}}\psi_{a,x+1}^{\alpha\dagger}U_{ab}(x)\psi^{\beta}_{b,x}\quad .
$$
The meson states are
\begin{eqnarray}
|S>&=&S|g.s.>\ ,\label{ssin}\\
|T_{\pm}>&=&T_{\pm}|g.s.>\ ,
\label{2tpm}\\
|T_{0}>&=&T_{0}|g.s.>\ ,
\label{to}
\end{eqnarray}
and are normalized as
\begin{eqnarray}
<S|S>&=&<g.s.|S^{\dagger}S|g.s.>=-\frac{4}{{\cal N}_{c}}<g.s.|H_{J}|g.s.>
\quad ,
\label{no1}\\
<T_{+}|T_{+}>&=&\langle g.s.|T_{+}^{\dagger}T_{+}|g.s.\rangle=\frac{2}{3}({\cal N}_{c}N+\frac{1}{{\cal N}_{c}}<g.s.|H_{J}|g.s.>)\quad .
\label{no2}
\end{eqnarray}
Furthermore,
\begin{equation}
<T_{0}|T_{0}>=<T_{-}|T_{-}>=<T_{+}|T_{+}> \ .
\label{no3}
\end{equation}
In Eqs.(\ref{no1},\ref{no2},\ref{no3}) $<g.s.|g.s.>=1$. 

The isosinglet energy, up to the second order in the strong coupling expansion, is $E_S=E_S^{(0)}+\epsilon^2E_S^{(2)}$ with 
\begin{eqnarray}
E_S^{(0)}&=&\frac{\langle S|H_{0}|S \rangle}{\langle S|S \rangle}=\frac{1}{2}\quad ,\\
E_S^{(2)}&=&\frac{\langle S|H^{\dagger}_h\Lambda_S H_h |S \rangle}{\langle S|S \rangle}\label{sotf}\quad .
\end{eqnarray}
In Eq.(\ref{sotf}) $\Lambda_S=\frac{\Pi_S}{E_S^{(0)}-H_{0}}$ and $\Pi_s$ 
is a projection operator orthogonal to $|S\rangle$. 
Using Eq.(\ref{crr}), one has
\begin{equation}
\Lambda_S\sum_{x=1}^{N}\sum_{\alpha,\beta=1}^2R_x^{(\alpha \alpha)}R_x^{(\beta \beta)}|g.s.\rangle=-\frac{1}{C_2^{f}({\cal N}_c)^2-1}
\sum_{x=1}^{N} \sum_{\alpha,\beta=1}^2
\left(C_2^{f}({\cal N}_c) R_x^{(\alpha \alpha)}R_x^{(\beta \beta)}+R_x^{(\alpha \beta)}R_x^{(\beta \alpha)}\right) |g.s.\rangle \quad .
\label{main1}
\end{equation}
As a result, Eq.(\ref{sotf}) may be written in terms of Heisenberg 
spin correlators; one has 
\begin{equation}
E_{S}^{(2)}=E_{g.s.}^{(2)}+4+\frac{\langle g.s.| -2\sum_{x=1}^N(\vec{S}_x \cdot \vec{S}_{x+2} -\frac{{\cal N}_c^2}{4}) +\frac{24}{{\cal N}_c^2} 
\sum_{x=1}^N\left[ (\vec{S}_x \cdot \vec{S}_{x+1})^2 -\frac{{\cal N}_c^4}{16} \right] |g.s. \rangle }{\langle g.s.|H_J|g.s. \rangle}\quad .
\label{se2}
\end{equation} 

At the zeroth order, the pseudoscalar triplet is degenerate with the pseudoscalar isosinglet $E_{T}^{(0)}=E_{S}^{(0)}=1/2$. 
Following the procedure previously outlined, one may compute the energy 
of the states (\ref{2tpm},\ref{to})) up to the second order in the strong 
coupling expansion. 
Using Eq.(\ref{main1}) written for 
$\Lambda_T=\frac{\Pi_T}{E_T^{(0)}-H_{0}}$, from Eq.(\ref{crl}), one gets
\begin{equation}
\Lambda_T\sum_{x=1}^{N}\sum_{\alpha,\beta=1}^2R_x^{(\alpha \alpha)}L_x^{(\beta \beta)}|g.s.\rangle=
-\frac{1}{C_{f}^{2}({\cal N}_c)} \sum_{x=1}^{N} \sum_{\alpha,\beta=1}^2 \left( R_x^{(\alpha \alpha)}L_x^{(\beta \beta)}
-\frac{1}{C_{f}^{2}({\cal N}_c)}\sum_{a,b=1}^{{\cal N}_c}\psi_{a x+1}^{\alpha \dagger}
\psi_{b x}^{\alpha}\psi_{b x}^{\beta \dagger}\psi_{a x+1}^{\beta}  \right) |g.s.\rangle \quad ,
\label{main2}
\end{equation}
the isotriplet energy $-$ to the second order in the strong coupling 
expansion $-$ is given by
\begin{equation}
E_{T}^{(2)}=E_{g.s.}^{(2)}+\frac{\langle g.s.| 4H_J-10\sum_{x=1}^{N}(\vec{S}_x\cdot \vec{S}_{x+2}-\frac{{\cal N}_c^2}{4}) 
+\frac{3}{4}{\cal N}_c^2 N |g.s. \rangle }
{{\cal N}_c^2N+\langle g.s.| H_J|g.s. \rangle}\quad .
\label{te2}
\end{equation}

In analogy with the Schwinger 
model~\cite{banks2, berruto1, berruto2}, there are also excitations created 
by acting on $|g.s.\rangle $ with the chiral currents. 
For the two-flavor continuum 't Hooft model, 
chiral current operators are given by
\begin{eqnarray}
j^5(x)&=&\overline{\psi}_{\alpha a}(x)\gamma^5\psi_{\alpha a}(x)\\
j^5_B(x)&=&\overline{\psi}_{\alpha a}(x)\gamma^5(\frac{\sigma_B}{2})_{\alpha \beta}\psi_{\beta a}(x)\quad .
\end{eqnarray}
The corresponding lattice operators at zero momentum are then
\begin{eqnarray}
S^5&=&R-L=\sum_{x=1}^Nj^5(x)\label{c51}\quad ,\\
T^5_+&=&(T^5_-)^{\dagger}=R^{(12)}-L^{(12)}=\sum_{x=1}^Nj^5_+(x)\label{c52}\quad ,\\
T^5_0&=&\frac{1}{\sqrt{2}}(R^{(11)}-L^{(11)}-R^{(22)}+L^{(22)})=\sum_{x=1}^Nj^5_3(x)\label{c53}\quad .
\end{eqnarray}
The meson states, created by (\ref{c51},\ref{c52},\ref{c53}) when acting on $|g.s.\rangle$, are
\begin{eqnarray}
|S^5\rangle&=&S^5|g.s.\rangle\quad ,\\
|T^5_{\pm}\rangle &=&T^5_{\pm}|g.s.\rangle \quad ,\\
|T^5_{0}\rangle &=&T^5_{0}|g.s.\rangle \quad .
\end{eqnarray}
They are normalized as
\begin{eqnarray}
\langle S^5|S^5 \rangle &=& \langle g.s.|S^{5\dagger}S^5|g.s.\rangle = -\frac{4}{{\cal N}_c}\langle g.s.|H_J|g.s.\rangle \quad ,
\label{5no1} \\
\langle T_+^5|T_+^5\rangle &=& \langle g.s.|T_{+}^{5\dagger}T_{+}^5|g.s.\rangle=\frac{2}{3}({\cal N}_{c}N+\frac{1}{{\cal N}_{c}}<g.s.|H_{J}|g.s.>)\quad .
\label{5no2}
\end{eqnarray}
Furthermore,
\begin{equation}
\langle T_0^5|T_0^5\rangle = \langle T_+^5|T_+^5\rangle = \langle T_-^5|T_-^5\rangle \quad .
\label{5no3}
\end{equation}

Following the same computational scheme used to 
find the masses of the states  $|S\rangle$ and $|T\rangle$, 
for the state $|S^5\rangle$ one gets
\begin{equation}
E_{S^5}^{(0)}=\frac{\langle S^5|H_{0}|S^5 \rangle}{\langle S^5|S^5 \rangle}=\frac{1}{2}\quad ,
\end{equation}
\begin{equation}
E_{S^5}^{(2)}=E_{g.s.}^{(2)}+12+\frac{\langle g.s.| -6\sum_{x=1}^N(\vec{S}_x \cdot \vec{S}_{x+2} -\frac{{\cal N}_c^2}{4}) +\frac{40}{{\cal N}_c^2} 
\sum_{x=1}^N\left[ (\vec{S}_x \cdot \vec{S}_{x+1})^2 -\frac{{\cal N}_c^4}{16} \right] |g.s. \rangle }{\langle g.s.|H_J|g.s. \rangle}\quad , 
\label{1es52}
\end{equation}
while, for the triplet $|T^5\rangle$, one gets
\begin{eqnarray}
E_{T^5}^{(0)}&=&\frac{1}{2}\quad ,\\
E_{T^5}^{(2)}&=&E_{g.s.}^{(2)}+\frac{\langle g.s.|-4H_J+2\sum_{x=1}^{N}(\vec{S}_x\cdot \vec{S}_{x+2}-\frac{{\cal N}_c^2}{4}) 
-\frac{11}{4}{\cal N}_c^2 N |g.s. \rangle }
{{\cal N}_c^2N+\langle g.s.| H_J|g.s. \rangle}\label{1et52}\quad .
\end{eqnarray}

As in the two-flavor Schwinger model~\cite{berruto2}, the strong coupling 
expansion of the meson masses can be evaluated in terms of spin-spin 
correlators on the ground state of the Heisenberg model.
Taking the planar limit ${\cal N}_c\longrightarrow \infty$, 
$S\longrightarrow \infty$, the quantum ground state $|g.s.\rangle$ 
becomes the classical N\'eel state $|N\rangle$ (\ref{neel}), and the 
Heisenberg spin-spin correlators can be easily evaluated  on $|N\rangle$ 
giving 
\begin{equation}
 \langle N|\vec{S}_x\cdot \vec{S}_{x+n}|N\rangle=\left\{\begin{array}{rl}
 \frac{{\cal N}_c^2}{4} &\mbox{if n is even}\\ 
 -\frac{{\cal N}_c^2}{4} &\mbox{if n is odd}\quad .
\end{array}
\right.
\end{equation}

The norms of the meson states given in 
Eqs.(\ref{no1},\ref{no2},\ref{no3},\ref{5no1},\ref{5no2},\ref{5no3}), 
for ${\cal N}_c\longrightarrow \infty$, are written as
\begin{eqnarray}
\langle S|S\rangle=\langle S^5|S^5\rangle &=&2{\cal N}_{c}N\quad ,\\
\langle T_{0}|T_{0}\rangle =\langle  T_{-}|T_{-}\rangle =\langle T_{+}|T_{+}\rangle
=\langle T^5_{0}|T^5_{0}\rangle =\langle  T^5_{-}|T^5_{-}\rangle =\langle T^5_{+}|T^5_{+}\rangle&=&\frac{{\cal N}_{c}}{3}N\quad ;
\end{eqnarray}
the ground state and meson energies read as 
\begin{eqnarray}
E_{g.s.}^{(0)}&=&0\label{pe}\quad ,\\
E_{g.s.}^{(2)}&=&-4{\cal N}_cN\quad ,
\end{eqnarray}
\begin{eqnarray}
E_{S}^{(0)}&=&\frac{1}{2}\quad ,\\
E_{S}^{(2)}&=&-4{\cal N}_cN+4\quad ,
\end{eqnarray}
\begin{eqnarray}
E_{T}^{(0)}&=&\frac{1}{2}\quad ,\\
E_{T}^{(2)}&=&-4{\cal N}_cN-\frac{5}{2}\quad ,
\end{eqnarray}
\begin{eqnarray}
E_{S^5}^{(0)}&=&\frac{1}{2}\quad ,\\
E_{S^5}^{(2)}&=&-4{\cal N}_cN+12\quad ,
\end{eqnarray}
\begin{eqnarray}
E_{T^5}^{(0)}&=&\frac{1}{2}\quad ,\\
E_{T^5}^{(2)}&=&-4{\cal N}_cN-\frac{3}{4}\label{ue}\quad .
\end{eqnarray}
As a result of Eqs.(\ref{pe}-\ref{ue}), to the second order in 
the strong coupling expansion, the meson masses are given by
\begin{eqnarray}
m_{S}&=&\frac{g_L^2a}{2}(E_S-E_{g.s.})=\frac{g_L^2a}{2}(\frac{1}{2}+4\epsilon^2)\quad ,\label{msin}\\
m_{T}&=&\frac{g_L^2a}{2}(E_T-E_{g.s.})=\frac{g_L^2a}{2}(\frac{1}{2}-\frac{5}{2}\epsilon^2)\quad ,\label{mtri}\\
m_{S^5}&=&\frac{g_L^2a}{2}(E_S^5-E_{g.s.})=\frac{g_L^2a}{2}(\frac{1}{2}+12\epsilon^2)\quad ,\label{5msin}\\
m_{T^5}&=&\frac{g_L^2a}{2}(E_T^5-E_{g.s.})=\frac{g_L^2a}{2}(\frac{1}{2}-\frac{3}{4}\epsilon^2)\quad .\label{5mtri}
\end{eqnarray}
 Again, in Eqs.(\ref{msin},\ref{mtri},\ref{5msin},\ref{5mtri})
one is left only with $N$-independent terms. This is a good check of our computation 
being the mass an intensive quantity. 

The two-flavor model spectrum exhibits $2^{{\cal N}_c}$ different kinds of baryons, which can be created at zero momentum by 
acting on the ground state with the color singlet operators $B^{\dagger}$
\begin{equation}
|B\rangle = B^{\dagger}|g.s.\rangle =\sum_{x=1}^{N}B^{\dagger}_x|g.s.\rangle =
\sum_{x=1}^{N}\psi_{1x}^{\alpha_1 \dagger}\psi_{2x}^{\alpha_2 \dagger}\ldots \psi_{{\cal N}_c x}^{\alpha_{{\cal N}_c} \dagger}|g.s. \rangle\quad ,
\end{equation}
where the greek indices $\alpha_1 \ldots \alpha_{{\cal N}_c}$ may take the values $1,2$. 
Again, we find that, in the strong coupling limit, the baryon masses 
are zero at zeroth order in the strong coupling expansion and acquire a mass proportional to ${\cal N}_c$ only at second order. Thus, also for 
${\cal N}_f=2$, baryons may be regarded as $QCD$ solitons, according to 
Witten's conjecture.

Let us consider, for example, the baryon built up with just flavor 1 quarks
\begin{equation}
|\Delta\rangle =\Delta^{\dagger}|g.s.\rangle =\sum_{x=1}^{N}\Delta^{\dagger}_x|g.s.\rangle  =
\sum_{x=1}^{N}\psi_{1x}^{1 \dagger}\psi_{2x}^{1 \dagger}\ldots \psi_{{\cal N}_c x}^{1\dagger} |g.s. \rangle\quad .
\end{equation}
At zeroth order in the strong coupling expansion, the baryon is massless, since the creation operator $\Delta^{\dagger}$ does not 
contain any color flux. One has then
\begin{equation}
H_0|\Delta\rangle =0 \quad .
\end{equation}
At the second order in the strong coupling expansion, since $H_h$ or $H_J$ do not commute with $\Delta^{\dagger}$, baryons acquire a mass given by
\begin{eqnarray}
E_{\Delta}^{(2)}&=&\frac{1}{\langle \Delta|\Delta \rangle}\langle \Delta|H_{h}^{\dagger}\frac{\Pi_{\Delta}}{E_{\Delta}^{(0)}-H_{0}}H_{h}|\Delta \rangle
=\frac{4}{{\cal N}_c}\left( \langle g.s.|H_J|g.s. \rangle +\frac{{\cal N}_c}{2}\right)=-4{\cal N}_{c}N+4\ ,\quad \ \\
m_{\Delta}^{(2)}&=&2g^2_La\quad .
\end{eqnarray}

\section{The chiral condensate}

In the continuum one-flavor 't Hooft model, the chiral symmetry is dynamically broken by the anomaly. The order parameter is the mass operator 
$M(x)=\overline{\psi}(x)\psi(x)$, which acquires a nonzero vacuum expectation value, giving rise to the chiral condensate~\cite{zhitnitsky}
\begin{equation}
\chi_{c}=\langle \overline{\psi}\psi \rangle=-{\cal N}_{c}(\frac{g^{2}{\cal N}_{c}}{12\pi})^{\frac{1}{2}}\quad .
\label{zchi2}
\end{equation} 

In this section we shall compute the lattice chiral condensate $\chi_L$ up to the fourth order in the strong coupling expansion. In the staggered fermion 
formalism the pertinent lattice operator is given by 
\begin{equation}
\sum_{a=1}^{{\cal N}_c} \overline{\psi}_a(x)\psi_a(x)\longrightarrow -\frac{(-1)^x}{a}(\sum_{a=1}^{{\cal N}_c}\psi^{\dagger}_{ax}\psi_{ax} -
\frac{{\cal N}_c}{2})\quad .
\end{equation}
The lattice chiral condensate may then be obtained using the mass 
operator 
\begin{equation}
M=-\frac{1}{Na}\sum_{x=1}^{N}\sum_{a=1}^{{\cal N}_c}(-1)^x \psi^{\dagger}_{ax}\psi_{ax}\quad ,
\label{mao}
\end{equation}
where the extra minus sign is put in Eq.(\ref{mao}) 
to give the same sign to the lattice and continuum chiral condensates. 

One has to evaluate the expectation value 
of Eq.(\ref{mao}) on the perturbed states $|p_{g.s.}\rangle$ generated by applying $H_{h}$ to $|g.s.\rangle$. One has 
\begin{equation}
|p_{g.s.}\rangle=|g.s.\rangle+\epsilon|p_{g.s.}^{(1)}\rangle+\epsilon^2 |p_{g.s.}^{(2)}\rangle \quad ,
\end{equation} 
where
\begin{eqnarray}
|p_{g.s.}^{(1)}\rangle&=&\frac{\Pi_{g.s.}}{E_{g.s.}^{(0)}-H_{0}}H_{h}|g.s.\rangle\quad ,\\
|p_{g.s.}^{(2)}\rangle&=&\frac{\Pi_{g.s.}}{E_{g.s.}^{(0)}-H_{0}}H_{h}\frac{\Pi_{g.s.}}{E_{g.s.}^{(0)}-H_{0}}H_{h}|g.s.\rangle\quad .
\end{eqnarray}
The lattice chiral condensate is then given by 
\begin{eqnarray}
\chi_{L}&=&\frac{\langle p_{g.s.}|M|p_{g.s.}\rangle}{\langle p_{g.s.}|p_{g.s.}\rangle}\nonumber\\
&=&\frac{\langle g.s.|M|g.s.\rangle+\epsilon^2 \langle p_{g.s.}^{(1)}|M|p_{g.s.}^{(1)}\rangle+\epsilon^4 \langle p_{g.s.}^{(2)}|M|p_{g.s.}^{(2)}\rangle}
{\langle g.s.|g.s.\rangle+\epsilon^2 \langle p_{g.s.}^{(1)}|p_{g.s.}^{(1)}\rangle+\epsilon^4 \langle p_{g.s.}^{(2)}|p_{g.s.}^{(2)}\rangle}\quad .
\label{1c}
\end{eqnarray}
The wave functions are normalized as
\begin{eqnarray}
\langle g.s.|g.s. \rangle &=& 1\quad ,\\
\langle p_{g.s.}^{(1)}|p_{g.s.}^{(1)}\rangle &=& 4{\cal N}_{c}N\quad ,\\
\langle p_{g.s.}^{(2)}|p_{g.s.}^{(2)}\rangle &=& 8{\cal N}_{c}^2N^2-8\frac{{\cal N}_{c}({\cal N}_{c}-2)}{{\cal N}_{c}-1}N\quad .
\end{eqnarray}
The expectation values of the mass operator $M$ are given by
\begin{eqnarray}
 \langle g.s.|M|g.s. \rangle &=&-\frac{1}{a}\frac{{\cal N}_{c}}{2}\quad ,\\
\langle p_{g.s.}^{(1)}|M|p_{g.s.}^{(1)}\rangle &=&-\frac{1}{a}(2{\cal N}_{c}^2N-8{\cal N}_{c})\quad ,\\
\langle p_{g.s.}^{(2)}|M|p_{g.s.}^{(2)}\rangle &=&-\frac{1}{a}(4{\cal N}_{c}^3N^2-\frac{36{\cal N}_{c}^3-40{\cal N}_{c}^2}{{\cal N}_{c}-1}N+\frac{32{\cal N}_{c}({\cal N}_{c}-2)}{{\cal N}_{c}-1})\quad .
\label{2m}
\end{eqnarray}
Using Eqs.(\ref{1c}-\ref{2m}), to the fourth order in $\epsilon$, 
one finds a nonvanishing chiral condensate for any finite ${\cal N}_c$. 
This is in agreement with the results of Ref.~\cite{grandou}. 
For ${\cal N}_c\longrightarrow \infty$ one gets 
\begin{equation}
\chi_{L}=-\frac{1}{a}{\cal N}_{c}(\frac{1}{2}-8\epsilon^2 +32\epsilon^4)\quad .
\label{chitho}
\end{equation}
We observe that
the lattice computation of the chiral condensate shows that, 
up to order $\epsilon^2$, Eq.(\ref{chitho}) is just ${\cal N}_c$ times 
the chiral condensate computed in Ref.~\cite{berruto1} for 
the one-flavor Schwinger model. 
Taking into account the contributions up to the 
fourth order in $\epsilon$, one has
\begin{equation}
\chi_{L}^{'t\ Hooft}={\cal N}_{c}(\chi_{L}^{Schwinger}+\frac{1}{a}64\epsilon^4)\quad .
\label{chichi}
\end{equation}
The difference is due to terms such as 
\begin{equation}
\langle g.s.|\sum_{x=1}^NR_xR_xL_xL_x|g.s.\rangle
\end{equation}
which are non vanishing only in the non-abelian model, since in the abelian
model the sites of $|g.s.>$ are either empty or occupied by just one particle:
$|g.s.>$ is annihilated when more than one hopping operator of the same type 
acts on the same site. The lattice 
computation, which brought us to Eq.(\ref{chichi}) shows that the 
the chiral condensate of the `t Hooft model factorizes to all orders 
in the product of ${\cal N}_c$ and a term whose numerical value is dominated 
by the contribution of the chiral condensate of the one flavor 
Schwinger model; as seen from Eq.(\ref{chichi}) the leading correction to the 
abelian chiral condensate is of order $O(\epsilon^4)$.
Eq.(\ref{chichi}) seems to suggest that the chiral condensate of the 
non-abelian model is essentially determined by the $U(1)$ abelian subgroup
of $U({\cal N}_c)$; the non abelian group contributing 
mainly a factor proportional to ${\cal N}_c$.
 If one puts ${\cal N}_c=1$ in Eq.(\ref{zchi2}) one gets an 
expression that differs less than 
the $2\%$ from the analytical expression of the Schwinger 
model chiral condensate~\cite{swieca}.

For ${\cal N}_f=2$, also on the lattice either the isoscalar $\langle \overline{\psi}\psi\rangle$ or the isovector 
$\langle \overline{\psi}\sigma^a\psi \rangle$ chiral condensates are zero to every order of perturbation theory due to the translational invariance 
of the quantum antiferromagnetic Heisenberg chain ground state. The proof 
is a straightforward generalization of the one given in \cite{berruto2} for the two-flavor Schwinger model. 

Just as in the two-flavor Schwinger model~\cite{berruto2}, 
the nonvanishing v.e.v. signaling the breaking of the 
$U_A(1)$ chiral symmetry is given by 
\begin{equation}
\langle F \rangle=\langle \overline{\psi^{2}}_L \overline{\psi^{1}}_L \psi_{R}^{1} \psi_R^{2} \rangle\quad .
\label{cf}
\end{equation}
To write this operator on the lattice one observes that
\begin{equation}
\sum_{a=1}^{{\cal N}_c}\overline{\psi^{\alpha}_{aL}}(x)\psi_{aR}^{\alpha}(x)\longrightarrow \frac{1}{2a}\frac{1}{2}
\left[ \sum_{a=1}^{{\cal N}_c}(
\psi^{\alpha \dagger}_{a x}\psi^{\alpha}_{a x}-\psi^{\alpha \dagger}_{a x+1}\psi^{\alpha}_{a x+1})+L^{\alpha}_x-R^{\alpha}_x\right]\quad .
\label{pp2}
\end{equation}
and, upon introducing the  occupation number 
operators $n_x^{\alpha}=\sum_{a=1}^{{\cal N}_c}\psi^{\alpha \dagger}_{a x}\psi^{\alpha}_{a x}$, one finds that, on the lattice, the operator $F$ is given by
\begin{equation}
F=-\frac{1}{16a^2 N} \sum_{x=1}^N \left\{ (n_x^1-n_{x+1}^1)(n_x^2-n_{x+1}^2)+(L_x^1-R_x^1)(L_x^2-R_x^2) \right\}\quad .
\end{equation}
The factor $1/2a$ in Eq.(\ref{pp2}) is due to the 
doubling of the lattice spacing in the antiferromagnetic bipartite lattice.

The strong coupling expansion carried up to the second order in $\epsilon=\frac{t}{g_L^2a^2}$, yields 
\begin{equation}
\langle F \rangle=\frac{\langle p_{g.s.}|F|p_{g.s.}\rangle}{\langle p_{g.s.}|p_{g.s.}\rangle}=
\frac{\langle g.s.|F|g.s.\rangle +\epsilon^2 \langle p_{g.s.}^{(1)}|F|p_{g.s.}^{(1)}\rangle }
{\langle g.s.|g.s.\rangle +\epsilon^2 \langle p_{g.s.}^{(1)}|p_{g.s.}^{(1)}\rangle }\quad .
\label{cc2}
\end{equation}
Since 
\begin{eqnarray}
\langle g.s.|g.s. \rangle &=& 1\quad , \\
\langle p_{g.s.}^{(1)} |p_{g.s.}^{(1)} \rangle &=&-\frac{16}{{\cal N}_c}\langle g.s.|H_J|g.s.\rangle\quad , 
\end{eqnarray}
and taking into account that  
\begin{eqnarray}
\langle g.s.|F|g.s.\rangle &=& -\frac{1}{16a^2N}\left(\frac{2}{3}\langle g.s.|H_J|g.s. \rangle -\frac{{\cal N}_c^2}{3}N\right)\quad ,\\
\langle p_{g.s.}^{(1)}|F|p_{g.s.}^{(1)} \rangle &=& -\frac{1}{16a^2N} \left[ \frac{32}{3{\cal N}_c}\langle g.s.|H_J^2|g.s. \rangle
+\frac{16}{3}{\cal N}_cN\langle g.s.|H_J|g.s. \rangle -\frac{16}{3}\langle g.s.|H_J|g.s. \rangle +8{\cal N}_c^2N  \right. \nonumber \\
&+&\left. \frac{16}{3}\langle g.s.|\sum_{x=1}^{N}(\vec{S}_x\cdot \vec{S}_{x+2}-
\frac{{\cal N}_c^2}{4})|g.s.\rangle +\frac{64}{3{\cal N}_c^2}\sum_{x=1}^{N}\langle g.s.|(\vec{S}_{x}\cdot \vec{S}_{x+1})^2|g.s.\rangle \right] \quad ,
\end{eqnarray}
in the planar limit ${\cal N}_c\rightarrow \infty$, from Eq.(\ref{cc2}),
 one gets
\begin{equation} 
\langle F \rangle=\frac{{\cal N}_c^2}{a^2}\left(0.042-0.750\epsilon^2 \right)
\quad .
\label{ccc2}
\end{equation}
The nonvanishing v.e.v. determined by Eq.(\ref{ccc2}) is the lattice relic of the $U_A(1)$ anomaly in the continuum theory. 
As evidenced in Ref.~\cite{berruto2} the operator $F$ describes 
on the lattice an umklapp process.
 
\section{Lattice versus continuum}    

It is our purpose in this section to compare the results 
of the strong coupling expansion with the continuum theory. 
To do this, one has to extrapolate the strong coupling series, 
derived under the assumption that the parameter $z=\epsilon^2=t^2/g^4a^4\ll 1$ to the region in which $z\gg 1$; this region corresponds to the 
continuum theory since $g^4a^4\longrightarrow 0$, $z\longrightarrow \infty$. To make this extrapolation possible, it is customary to make 
use of Pad\'e approximants, which allow one to extrapolate a series 
expansion beyond the convergence radius. 

Let us consider the one-flavor model. We first compute the lattice light velocity by equating the lattice chiral condensate 
given in Eq.(\ref{chitho}) to its continuum counterpart 
written in Eq.(\ref{zchi2})
\begin{equation}
-\frac{1}{a}{\cal N}_{c}(\frac{1}{2}-8z+32z^2)=-{\cal N}_{c}(\frac{g^{2}{\cal N}_{c}}{12\pi})^{\frac{1}{2}}\quad .
\label{rcp}
\end{equation}
Eq.(\ref{rcp}) is true only when Pad\'e approximants are used since, as it stands, the left hand side holds only for $z\ll 1$, while the right 
hand side provides the value of the chiral condensate to be obtained when $z\approx \infty$. Using
\begin{equation}
\frac{1}{a}=(\frac{g^2{\cal N}_c}{t})^{\frac{1}{2}} z^{\frac{1}{4}}\quad ,
\label{r1a}
\end{equation}
Eq.(\ref{rcp}) reads 
\begin{equation}
(\frac{g^2{\cal N}_c}{t})^{\frac{1}{2}} z^{\frac{1}{4}}(\frac{1}{2}-8z+32z^2) =(\frac{g^{2}{\cal N}_{c}}{12\pi})^{\frac{1}{2}}\quad .
\label{rcp2}
\end{equation}
Due to the factor $z^{\frac{1}{4}}$, one takes the eighth power of Eq.(\ref{rcp2}), then expands the left hand side polynomial 
of Eq.(\ref{rcp2}) for small $z$ and finally 
takes the continuum limit $z\longrightarrow \infty$; one gets then the 
lattice light velocity
\begin{equation}
t=\frac{12\pi}{16(34)^{\frac{1}{4}}}=0.9757
\end{equation}
which lies $2.4\%$ below the exact value. 

If one extrapolates the mass of the pseudoscalar excitation 
Eq.(\ref{pse}) to the continuum, one obtains
\begin{equation}
m_{P}=\frac{g^2a}{2}(\frac{1}{2}-4z+80z^2)\quad .
\label{mp2}
\end{equation}
Using 
\begin{equation}
\frac{ga}{t}=\frac{1}{z^{\frac{1}{4}}}\quad ,
\label{gat}
\end{equation}
taking the eighth power of Eq.(\ref{mp2}), expanding for small $z$ and 
then taking the limit $z\longrightarrow \infty$, one gets
\begin{equation}
\frac{m_{P}}{g}=0.6655\quad ,
\label{nmpse}
\end{equation}
which agrees within 16\% with the result obtained in~\cite{bhattacharya} in the continuum for ${\cal N}_c=3$ and with the lattice numerical calculation 
of Ref.~\cite{luo}. 
If one extrapolates to the continuum the scalar meson mass 
given in Eq.(\ref{msca}), at this order in the strong coupling expansion, 
one finds that it is degenerate with the pseudoscalar excitation mass 
given by Eq.(\ref{nmpse}).

Let us now turn our attention to the two-flavor model. 
It would be desiderable to extrapolate to the continuum the v.e.v. 
given in (\ref{ccc2}). Unfortunately, to our knowledge, 
no one computed Eq.(\ref{cf}) in the continuum. Thus, Eq.(\ref{ccc2}) cannot 
be used to fix the lattice light velocity, which we assume to be 
equal to 1 in the subsequent computation. 
Using Eq.(\ref{r1a}), Eq.(\ref{ccc2}) reads
\begin{equation}
\langle F\rangle=g^2{\cal N}_c^3z^{\frac{1}{2}}(0.042-0.750z)\quad .
\label{Fe}
\end{equation}
Due to the factor $z^{\frac{1}{2}}$, one squares Eq.(\ref{Fe}), 
then expands (\ref{Fe}) for small $z$ and 
takes the continuum limit 
$z\longrightarrow \infty$, getting 
\begin{equation}
\langle F\rangle=0.007g^2{\cal N}_c^3\quad .  
\end{equation}

The mass of the isosinglet excitation 
given in Eq.(\ref{msin}), due to (\ref{gat}), reads
\begin{equation}
\frac{m_S}{g}=\frac{1}{z^{\frac{1}{4}}}(\frac{1}{4}+2z)\quad .
\label{msinr}
\end{equation}
Again, the factor $z^{\frac{1}{4}}$ leads one to take the fourth power 
of Eq.(\ref{msinr}), to extrapolate the polynomial for small $z$ 
and finally take the continuum limit $z\longrightarrow \infty$, getting
\begin{equation}
\frac{m_S}{g}=0.5946\quad .
\label{nmsin}
\end{equation}
Following the same procedure used to compute $m_S$, one gets the mass of 
the singlet excitation Eq.(\ref{5msin})
\begin{equation}
\frac{m_{S^5}}{g}=0.7825\quad .
\label{5nmsin}
\end{equation}
To our knowledge, no one computed the isosinglet masses 
given in Eqs.(\ref{nmsin},\ref{5nmsin}) in the continuum theory. 
According to our previous 
experience~\cite{berruto1,berruto2} of strong coupling computations, 
we believe that Eqs.(\ref{nmsin},\ref{5nmsin}) give a good numerical esteem of their continuum values.
The procedure used to compute $m_S$ and $m_{S^{5}}$ doesn't work for the isotriplet masses given in Eqs.(\ref{mtri},\ref{5mtri}), since they have a 
second order negative correction. In order to extrapolate the isotriplet masses Eqs.(\ref{mtri},\ref{5mtri}) to the continuum, one should 
compute their strong coupling expansions to higher orders;
 a straightforward, but algebraically involved, computation 
which is beyond the scope of this paper.

\section{Concluding remarks}

In this paper we focused on the strong coupling analysis 
of the spectrum and chiral symmetry breaking 
pattern of $QCD_2$ with one and two fermion flavors. 
For both models, the vacuum, in the infinite coupling limit, 
is given by the ground states of the Ising and Heisenberg antiferromagnets, 
respectively: 
while the ground state of a one dimensional Ising antiferromagnet is a well defined classical N\'eel configuration, the ground state of a quantum 
spin-${\cal N}_c/2$ Heisenberg antiferromagnetic chain is a highly non trivial isotropic and translationally invariant singlet configuration. 
For the computation of both the spectrum and the chiral condensate of the gauge model, one needs only to know the symmetries of  the quantum 
ground state; these symmetries allow to express the relevant physical 
quantities only in terms of v.e.v.'s of spin correlators. 
For the two flavor model $-$ for ${\cal N}_c\longrightarrow \infty$ 
($S\longrightarrow \infty$) $-$ one needs to evaluate spin correlators 
only on the N\'eel classical configuration. This is 
one of the remarkable simplifications introduced by taking the 
large-${\cal N}_c$ limit in lattice $QCD_2$.

Mesons are created by applying to the vacuum lattice currents generating 
quark trasport. Baryons are, instead, local disturbances 
created by applying to the vacuum suitable color singlet operators. 
Baryons are massless in the infinite coupling limit, while they start 
acquiring a mass already at the second order in the strong 
coupling expansion; this is in agreement with the Witten 
conjecture~\cite{witten} that the baryons are the solitons of $QCD$. 
For what concerns the meson spectrum, one has that, in the one-flavor model, 
the extrapolation to the continuum of 
the lowest lying meson is in good agreement with the continuum result of Ref.~\cite{bhattacharya} and with the lattice results of Ref.~\cite{luo}.

In the one-flavor model we showed that the discrete axial symmetry 
is spontaneously broken by the vacuum and we computed the lattice 
chiral condensate $\langle \bar{\psi}\psi\rangle$ up to the fourth 
order in the strong coupling expansion. The extrapolation to the continuum 
of the lattice chiral condensate agrees within $2.4\%$ with the 
result of Ref.~\cite{zhitnitsky}. In the two-flavor model, 
the pertinent non-zero chiral condensate is 
$\langle F \rangle=\langle \overline{\psi^{2}}_L \overline{\psi^{1}}_L 
\psi_{R}^{1} \psi_R^{2} \rangle$, which 
is the lattice relic of the $U_A(1)$ anomaly in the continuum theory. 
We computed $\langle F \rangle$ up 
to the second order in the strong coupling expansion.

The strong coupling lattice computation of the chiral condensate 
for the one-flavor lattice $U({\cal N}_c)$ 't Hooft model shows 
that the numerical value of the chiral condensate is determined 
mainly by the $U(1)$ abelian subgroup of $U({\cal N}_c)$, 
the non-abelian 
group contributing essentially with a multiplicative factor 
${\cal N}_c$. 
To the second  order in the strong coupling expansion, the 
chiral condensate is just proportional to the one computed 
in Ref.\cite{berruto1} for the one-flavor Schwinger model. One may 
be led to conjecture 
that this result holds also for the continuum $U({\cal N}_c)$ 't Hooft model.

We feel that the relation between gauge models and antiferromagnetic spin 
systems in the strong coupling limit is particularly useful, 
since it provides an intuitive representation of 
the ground state and the excitations of the gauge theory
and, in principle, it enables one 
to explore the phase transition of a gauge theory 
using the knowledge of the phases of magnetic systems. 
It would be very interesting to investigate if this correspondence
between gauge theories and quantum 
antiferromagnets survives also in the weak coupling limit. 
An interesting proposal in this direction has been made recently by 
Weinstein in Ref.~\cite{weinstein2}, where, using the 
Contractor Renormalization Group method, he 
established the equivalence of various Hamiltonian free fermion 
theories with a class of generalized frustrated antiferromagnets.     
\vskip 0.3truein  
\noindent
{\large \bf Acknowledgements}
\vskip 0.1truein
\noindent
We thank Prof. G. W. Semenoff and Dr. E. Coletti for many valuable 
discussions. This work has been supported by grants from the Istituto Nazionale di Fisica Nucleare and M.U.R.S.T. .   

\section{Appendix A}

In this appendix we provide a table of the integrals over the group elements of $SU({\cal N}_{c})$ useful $-$ for large ${\cal N}_c$ $-$ 
in the strong coupling expansions up to the fourth order in $\epsilon$. 
It is well known that a basic ingredient to formulate $QCD$ on a lattice is to define the measure of integration over the gauge degrees of freedom. 
Unlike the continuum gauge fields, the lattice gauge fields are $SU({\cal N}_{c})$ matrices with elements bounded in the range $\left[0,1\right]$; 
Wilson \cite{wilson} proposed an invariant group measure, the Haar measure, for the integration over the group elements. The integral is defined so that, for any 
elements $g_1$ and $g_2$ of the group, one has
\begin{equation}
\int dU\ f(U)= \int dU\ f(Ug_1)=\int dU\ f(g_2U)\quad ,
\label{dud}
\end{equation}
with $f(U)$ a generic function over the group. 
When used in non-perturbative studies of gauge theory, 
the definition (\ref{dud}) avoids the problem of having to introduce a gauge 
fixing, since the field variables are compact. 
The measure is normalized as 
\begin{equation}
\int dU =1\quad .
\end{equation}
The strong coupling expansion for an $SU({\cal N}_c)$ gauge theory depends 
on the following identities for integration over link 
matrices \cite{creutz}
\begin{eqnarray}
\int dU\ U_{ab}&=&\int dU\ U_{ab}^{\dagger}=0\quad ,\label{i1}\\
\int dU\ U_{ab}U_{cd}^{\dagger}&=&\frac{1}{{\cal N}_{c}}\delta_{ab}\delta_{cd}\quad ,\label{i2}\\ 
\int dU\ U_{ab}U_{cd}^{\dagger}U_{ef}U_{gh}^{\dagger}&=&\frac{1}{{\cal N}_{c}^2-1}(\delta_{ad}\delta_{bc}\delta_{eh}\delta_{fg}+\delta_{ah}\delta_{bg}
\delta_{cf}\delta_{de})\nonumber\\
& &-\frac{1}{{\cal N}_{c}({\cal N}_{c}^2-1)}(\delta_{ad}\delta_{bg}\delta_{eh}\delta_{fc}+\delta_{ah}\delta_{bc}\delta_{ed}
\delta_{fg})\quad .\label{i3}
\end{eqnarray} 
Group integration gives a non-zero result only if each link exhibits a combination of matrices, from which a color singlet can be formed. Using 
the computational scheme pionered by Creutz \cite{creutz}, we needed to compute the group integral over six elements, which occurs at the fourth 
order in the mass spectrum strong coupling expansion. The pertinent integral is given by
\begin{eqnarray}
\int dU\ U_{ab}U_{cd}^{\dagger}U_{ef}U_{gh}^{\dagger}U_{il}U_{mn}^{\dagger}&=&\frac{{\cal N}^{2}_c-2}{{\cal N}_c({\cal N}_c^2-1)({\cal N}_c^2-4)}\quad
(\delta_{ad}\delta_{eh}\delta_{in}\delta_{bc}\delta_{fg}\delta_{lm}+\delta_{ad}\delta_{en}\delta_{hi}\delta_{bc}\delta_{fm}\delta_{gl}\nonumber\\
& &\quad\quad\quad\quad\quad\quad\quad\quad\quad\quad
+\delta_{ah}\delta_{de}\delta_{in}\delta_{bg}\delta_{cf}\delta_{lm}+\delta_{an}\delta_{di}\delta_{eh}\delta_{bm}\delta_{cl}\delta_{fg}\nonumber\\
& &\quad\quad\quad\quad\quad\quad\quad\quad\quad\quad
+\delta_{an}\delta_{de}\delta_{hi}\delta_{bm}\delta_{cf}\delta_{gl}+\delta_{ah}\delta_{di}\delta_{en}\delta_{bg}\delta_{cl}\delta_{fm})\nonumber\\
& &-\frac{1}{({\cal N}_c^2-1)({\cal N}_c^2-4)}\quad 
(\delta_{ad}\delta_{en}\delta_{hi}\delta_{bc}\delta_{fg}\delta_{lm}+\delta_{ah}\delta_{de}\delta_{in}\delta_{bc}\delta_{fg}\delta_{lm}\nonumber\\
& &\quad\quad\quad\quad\quad\quad\quad\quad\quad\quad+
\delta_{an}\delta_{di}\delta_{eh}\delta_{bc}\delta_{fg}\delta_{lm}+\delta_{ad}\delta_{eh}\delta_{in}\delta_{bc}\delta_{fm}\delta_{gl}\nonumber\\
& &\quad\quad\quad\quad\quad\quad\quad\quad\quad\quad+
\delta_{an}\delta_{de}\delta_{hi}\delta_{bc}\delta_{fm}\delta_{gl}+\delta_{ah}\delta_{di}\delta_{en}\delta_{bc}\delta_{fm}\delta_{gl}\nonumber\\
& &\quad\quad\quad\quad\quad\quad\quad\quad\quad\quad+
\delta_{ad}\delta_{eh}\delta_{in}\delta_{bg}\delta_{cf}\delta_{lm}+\delta_{an}\delta_{de}\delta_{hi}\delta_{bg}\delta_{cf}\delta_{lm}\nonumber\\
& &\quad\quad\quad\quad\quad\quad\quad\quad\quad\quad+
\delta_{ah}\delta_{di}\delta_{en}\delta_{bg}\delta_{cf}\delta_{lm}+\delta_{ad}\delta_{eh}\delta_{in}\delta_{bm}\delta_{cl}\delta_{fg}\nonumber\\
& &\quad\quad\quad\quad\quad\quad\quad\quad\quad\quad+
\delta_{an}\delta_{de}\delta_{hi}\delta_{bm}\delta_{cl}\delta_{fg}+\delta_{ah}\delta_{di}\delta_{en}\delta_{bm}\delta_{cl}\delta_{fg}\nonumber\\
& &\quad\quad\quad\quad\quad\quad\quad\quad\quad\quad+
\delta_{ad}\delta_{en}\delta_{hi}\delta_{bm}\delta_{cf}\delta_{gl}+\delta_{ah}\delta_{de}\delta_{in}\delta_{bm}\delta_{cf}\delta_{gl}\nonumber\\
& &\quad\quad\quad\quad\quad\quad\quad\quad\quad\quad+
\delta_{an}\delta_{di}\delta_{eh}\delta_{bm}\delta_{cf}\delta_{gl}+\delta_{ad}\delta_{en}\delta_{hi}\delta_{bg}\delta_{cl}\delta_{fm}\nonumber\\
& &\quad\quad\quad\quad\quad\quad\quad\quad\quad\quad+
\delta_{ah}\delta_{de}\delta_{in}\delta_{bg}\delta_{cl}\delta_{fm}+\delta_{an}\delta_{di}\delta_{eh}\delta_{bg}\delta_{cl}\delta_{fm})\nonumber\\
& &+\frac{2}{{\cal N}_c({\cal N}_c^2-1)({\cal N}_c^2-4)}\quad 
(\delta_{an}\delta_{de}\delta_{hi}\delta_{bc}\delta_{fg}\delta_{lm}+\delta_{ah}\delta_{di}\delta_{en}\delta_{bc}\delta_{fg}\delta_{lm}\nonumber\\
& &\quad\quad\quad\quad\quad\quad\quad\quad\quad\quad+
\delta_{ah}\delta_{de}\delta_{in}\delta_{bc}\delta_{fm}\delta_{gl}+\delta_{an}\delta_{di}\delta_{eh}\delta_{bc}\delta_{fm}\delta_{gl}\nonumber\\
& &\quad\quad\quad\quad\quad\quad\quad\quad\quad\quad+
\delta_{ad}\delta_{en}\delta_{hi}\delta_{bg}\delta_{cf}\delta_{lm}+\delta_{an}\delta_{di}\delta_{eh}\delta_{bg}\delta_{cf}\delta_{lm}\nonumber\\
& &\quad\quad\quad\quad\quad\quad\quad\quad\quad\quad+
\delta_{ad}\delta_{en}\delta_{hi}\delta_{bm}\delta{cl}\delta_{fg}+\delta_{ah}\delta_{de}\delta_{in}\delta_{bm}\delta_{cl}\delta_{fg}\nonumber\\
& &\quad\quad\quad\quad\quad\quad\quad\quad\quad\quad+
\delta_{ad}\delta_{eh}\delta_{in}\delta_{bm}\delta_{cf}\delta_{gl}+\delta_{ah}\delta_{di}\delta_{en}\delta_{bm}\delta_{cf}\delta_{gl}\nonumber\\
& &\quad\quad\quad\quad\quad\quad\quad\quad\quad\quad+
\delta_{ad}\delta_{eh}\delta_{in}\delta_{bg}\delta_{cl}\delta_{fm}+\delta_{an}\delta_{de}\delta_{hi}\delta_{bg}\delta_{cl}\delta_{fm})\quad .\nonumber\\
& &\label{i4}
\end{eqnarray}
 
\section{Appendix B}
In this appendix we report technical details useful to determine the 
$QCD_2$ hadron spectrum and chiral condensate of sections 4 and 5. 
It is very useful to obtain the inverse of Eq.(\ref{scsp}); this is given by
\begin{equation}
\sum_{a=1}^{{\cal N}_c}\psi_{ax}^{\alpha \dagger}\psi_{ax}^{\beta}=\vec{S}_{x}\cdot \vec{\sigma}_{\beta \alpha}+{\cal N}_c\frac{\delta^{\alpha \beta}}{2}
\quad ,
\end{equation}
where the Pauli matrices satisfy 
\begin{eqnarray}
tr(\sigma^a\sigma^b)&=&2\delta^{ab}\quad ,\\
tr(\sigma^a\sigma^b\sigma^c)&=&2i\epsilon^{abc}\quad ,\\
tr(\sigma^a\sigma^b\sigma^c\sigma^d)&=&2\delta^{ab}\delta^{cd}-2\delta^{ac}\delta^{bd}+2\delta^{ad}\delta^{bc}\quad .
\end{eqnarray}

In order to explicitly compute Eqs.(\ref{egs4},\ref{ep2},\ref{ep4},\ref{es2},\ref{es4}) and Eqs.(\ref{main1},\ref{main2},
\ref{1es52},\ref{1et52}), one needs the following comutators, which hold effectively on $|g.s.\rangle$; namely, 
\begin{eqnarray}
\left[H_0,R^{(\alpha \alpha)}R^{(\beta \beta)}\right]&=&2C_2^{f}({\cal N}_c)R^{(\alpha \alpha)}R^{(\beta \beta)}-\sum_{x=1}^{N}
R_x^{(\alpha \alpha)}R_x^{(\beta \beta)}\quad ,\label{crr}\\
\left[H_0,L^{(\alpha \alpha)}L^{(\beta \beta)}\right]&=&2C_2^{f}({\cal N}_c)L^{(\alpha \alpha)}L^{(\beta \beta)}-\sum_{x=1}^{N}
L_x^{(\alpha \alpha)}L_x^{(\beta \beta)}\quad ,\label{cll}\\
\left[H_0,R^{(\alpha \alpha)}L^{(\beta \beta)}\right]&=&2C_2^{f}({\cal N}_c)R^{(\alpha \alpha)}L^{(\beta \beta)}-\sum_{x=1}^{N}
\psi_{ax+1}^{\alpha \dagger}\psi_{bx}^{\alpha}\psi_{bx}^{\beta \dagger}\psi_{ax+1}^{\beta}\quad ,\label{crl}
\end{eqnarray}  
where $\alpha=\beta=1$ in the one-flavor model.

In section 4 and 5 we used also the commutators
\begin{eqnarray}
\left[ R^{(\alpha \alpha)}_x,L^{(\beta \beta)}_y\right]&=&(n_{x+1}-n_x)\delta^{\alpha \beta}\delta_{xy}\quad ,\\
\left[R^{(\alpha \alpha)}_x,R^{(\beta \beta)}_y\right]&=&(\psi_{ax+1}^{\alpha \dagger}U_{ab}(x)U_{bc}(x-1)\psi_{cx-1}^{\alpha}
\delta_{x,y+1}-\psi_{ax+2}^{\alpha \dagger}U_{ab}(x+1)U_{bc}(x)\psi_{cx}^{\alpha}\delta_{y,x+1})\delta^{\alpha \beta}\quad ,\quad  \quad \quad \quad \\
\left[L^{(\alpha \alpha)}_x,L^{(\beta \beta)}_y\right]&=&(\psi_{ax}^{\alpha \dagger}U^{\dagger}_{ab}(x)U^{\dagger}_{bc}(x+1)\psi_{cx+2}^{\alpha}
\delta_{y,x+1}-\psi_{ax-1}^{\alpha \dagger}U^{\dagger}_{ab}(x-1)U^{\dagger}_{bc}(x)\psi_{cx+1}^{\alpha}\delta_{x,y+1})\delta^{\alpha \beta}\quad .\quad 
\quad \quad \quad
\end{eqnarray}

\end{document}